\documentclass[trackchanges,twocolumn]{aastex7}

\usepackage{makecell}

\newcommand{\todo}[1]{\textcolor{black}{#1}}

\usepackage[acronym,nohypertypes={acronym}]{glossaries} 
\newacronym{LSST}{LSST}{Legacy Survey of Space and Time}
\newacronym{JWST}{JWST}{James Webb Space Telescope}
\newacronym{HST}{HST}{Hubble Space Telescope}
\newacronym{FWHM}{FWHM}{full width at half maximum}
\newacronym{PSF}{PSF}{point spread function}
\newacronym{ATLAS}{ATLAS}{Asteroid Terrestrial-impact Last Alert System}
\newacronym{NEA}{NEA}{near-Earth
asteroids}
\newacronym{ATClean}{\texttt{ATClean}}{ATLAS Clean}
\newacronym{SNR}{SNR}{signal-to-noise ratio}
\newacronym{RSG}{RSG}{red supergiant}
\newacronym{SN}{SN}{supernova}
\newacronym{SN II}{SN II}{Type II supernova}
\newacronym{SN Type Ia}{SN Type Ia}{Type Ia supernova}
\newacronym{MIR}{MIR}{mid-infrared}
\newacronym{SED}{SED}{spectral energy distribution}
\newacronym{CSM}{CSM}{circumstellar material}
\newacronym{ALMA}{ALMA}{Atacama Large Millimeter/submillimeter Array}
\newacronym{GOTO}{GOTO}{Gravitational-wave Optical Transient Observer}
\newacronym{WFPC2}{WFPC2}{Wide Field Planetary Camera 2}
\newacronym{DECam}{DECam}{Dark Energy Camera}
\newacronym{ZTF}{ZTF}{Zwicky Transient Facility}
\newacronym{PTF}{PTF}{Palomar Transient Factory}
\newacronym{ACS}{ACS}{Advanced Camera for Surveys}
\newacronym{WFC}{WFC}{Wide Field Channel} 
\newacronym{CHIPS}{CHIPS}{Complete History of Interaction-Powered Supernovae}
\newacronym{DETECT}{\texttt{DETECT}}{Detection Efficiency and Threshold Estimation for Characterization of Transients}
\newacronym{DR1}{DR1}{Data Release 1}
\newacronym{DP2}{DP2}{Data Preview 2}
\newacronym{DP1}{DP1}{Data Preview 1}
\newacronym{DP0}{DP0}{Data Preview 0}
\newacronym{YSE}{YSE}{Young Supernova Experiment}
\newacronym{SDSS}{SDSS}{Sloan Digital Sky Survey}
\newacronym{TNS}{TNS}{Transient Name Server}

\newacronym{LSSTComCam}{LSSTComCam}{LSST Commissioning Camera}
\newacronym{LSSTCam}{LSSTCam}{LSST Camera}
\newacronym{RSP}{RSP}{Rubin Science Platform}
\newacronym{DESC}{DESC}{Dark Energy Science Collaboration}
\newacronym{DC2}{DC2}{Data Challenge 2}
\newacronym{IAU}{IAU}{International Astronomical Union}
\newacronym{TDE}{TDE}{tidal disruption event}
\newacronym{AGN}{AGN}{active galactic nucleus}
\newacronym{LBV}{LBV}{luminous blue variable}

\graphicspath{{./}{figures/}}

\begin{document}


\title{DETECT: A Pipeline to Quantify Detection Thresholds in Rubin for Nearby Targets Embedded in Bright Host Galaxies}

\correspondingauthor{Tobias G\'eron}
\email{tobias.geron@utoronto.ca}

\author[0000-0002-6851-9613]{Tobias G\'eron}
\affiliation{Dunlap Institute for Astronomy \& Astrophysics, University of Toronto, 50 St. George Street, Toronto, ON M5S 3H4, Canada}
\email{tobias.geron@utoronto.ca}

\author[0000-0001-7081-0082]{Maria R. Drout}
\affiliation{David A. Dunlap Department of Astronomy \& Astrophysics, University of Toronto, 50 St. George Street, Toronto, ON, M5S 3H4, Canada}
\email{-}

\author[0000-0002-3934-2644]{W.~V.~Jacobson-Gal\'{a}n}
\altaffiliation{NASA Hubble Fellow}
\affiliation{Cahill Center for Astrophysics, California Institute of Technology, MC 249-17, 1216 E California Boulevard, Pasadena, CA, 91125, USA}
\email{-}

\author[0000-0002-5740-7747]{C. D. Kilpatrick}
\affiliation{Center for Interdisciplinary Exploration and Research in Astrophysics (CIERA) and Department of Physics and Astronomy, Northwestern University, Evanston, IL 60208, USA}
\email{-}


\begin{abstract}
The final stages of stellar evolution can be constrained by studying pre-SN variability. The incredible amount of data coming from the upcoming Rubin Legacy Survey of Space and Time (LSST) will be fundamental to this type of work. However, robustly measuring pre-SN variability can be hard, as even state-of-the-art image subtraction pipelines struggle when the target is embedded in a bright nearby galaxy. We developed Detection Efficiency and Threshold Estimation for Characterization of Transients (\texttt{DETECT}) to tackle this problem. It performs a series of source injection, image subtraction, and forced photometry to obtain reliable detection thresholds tailored to a specific location within a given host galaxy. We first validate the pipeline using simulated data from Rubin DP0 and then apply it to a sample of \todo{15} targets found in Rubin DP1. We demonstrate that \texttt{DETECT} is capable of identifying pre-SN variability while calculating reliable upper limits and suppressing false positives for targets embedded in bright host galaxies. Most of the false positives in this work occurred when the signal-to-noise ratio (SNR) was between 5 and 10, while no false positives were found when the SNR was greater than 10. Finally, even though \texttt{DETECT} was originally developed in the context of pre-SN variability, it is broadly applicable to any situation where detections are uncertain and robust upper limits are needed. 




\end{abstract}












\section{Introduction}



Traditionally, massive stars were thought to be relatively static in their final years prior to core-collapse. However, recent observations with modern surveys have challenged this idea. For example, a subset of hydrogen-rich \gls{SN II}, called SN Type IIn, contain ``narrow'' emission lines. These narrow emission lines result from the interaction between the SN ejecta and the dense, slowly moving pre-existing \gls{CSM} \citep{gal_yam_2017, smith_2017}. While the origin of the \gls{CSM} and nature of the progenitors of SN Type IIn are debated, one mechanism that is often invoked to explain the presence of this dense \gls{CSM} is enhanced mass loss due to eruptive outbursts prior to explosion (e.g., \citealt{dessart_2010, fuller_2017, tsang_2022}). Such outbursts would appear in the pre-explosion light curve as short ``bumps''. This type of precursor emission has been observed for many Type IIn, including SN 2009ip \citep{pastorello_2013}, 2010mc \citep{ofek_2013}, 2011ht \citep{fraser_2013}, and 2023vbg \citep{goto_2025}. In fact, precursor emission seems to be relatively common for SN Type IIn. For example, \citet{strotjohann_2021} find that $25^{+44}_{-20}$\% of Type IIn show a month-long precursor emission event brighter than $-13$ mag in the $r$-band within three months before the explosion, while \citet{ofek_2014} find that more than half of SN Type IIn have at least one pre-explosion outburst brighter than $3 \times 10^{7} L_\odot$ within $\sim$0.33 years prior to explosion. Fainter precursor emission may be even more common, but could often be missed due to limited survey depth and cadence.

Furthermore, as it has become more common for SN to be discovered within days (or hours) of explosion, there is an increasing body of evidence that some amount of dense \gls{CSM} may also be present close to the progenitors of many otherwise `normal' classes of SN. For example, the spectra of some SN Type II also show transient narrow emission lines, reminiscent of the aforementioned SN Type IIn. However, unlike SN Type IIn, these features are short-lived, typically lasting only hours to days, before the rapidly expanding ejecta overtake the slower \gls{CSM} and Doppler-broadened ejecta lines dominate the spectra \citep{dessart_2017, yaron_2017, jacobson_galan_2023, jacobson_galan_2024b}. These SN are called ``IIn-like''. Interestingly, these SN Type II with IIn-like features appear to be relatively common: more than 36\% of SN Type II observed within two days of explosion display these IIn-like features \citep{bruch_2021, bruch_2023}, although the short nature of these emission lines makes them difficult to detect.

As noted above, precursor emission appears to be prevalent among SN Type IIn. However, it is unclear how common precursor emission is among these SN Type II with IIn-like features. The short-lived nature of the narrow emission lines for SN Type II with IIn-like features implies that their surrounding \gls{CSM} is possibly less massive. This, in turn, suggests that the corresponding original precursor emission is likely to be fainter and harder to observe. While \citet{jacobson_galan_2022} did find evidence for precursor emission in SN 2020tlf, this was not observed for SN 2023ixf \citep{dong_2023, ransome_2024,rest_2025} or SN 2024ggi (\citealp{shrestha_2024}; T. G{\'e}ron et al. in prep.), despite their proximity. SN 2023ixf and SN 2024ggi were located at a distance of 6.85 and 7.2 Mpc, respectively, and precursor emission brighter absolute magnitudes of -8 to -11 was ruled out for both cases, which is deeper than typical detected precursor emission (\citealp{rest_2025}; T. G{\'e}ron et al. in prep.). Additionally, \citet{johnson_2018} finds no clear evidence for variability in the pre-explosion light curves in their sample of four nearby SN Type IIP/IIL. Thus, it is still an open question whether SN Type IIP/IIL and SN Type II with IIn-like features exhibit significant precursor emission. If they do not, then other mechanisms must be invoked to explain the \gls{CSM} surrounding the SN Type II with IIn-like features. A few other mechanisms have been proposed, such as binary interactions \citep{matsuoka_2024}, extended atmospheres/chromospheres \citep{dessart_2017, fuller_2024}, and `superwinds' \citep{yoon_2010}. 


The key to understanding the origin of this dense \gls{CSM} is to study the precursor emission (or lack thereof) for a large sample of SN. This will be possible with the upcoming Rubin \gls{LSST} \citep{ivezic_2019}. The Rubin \gls{LSST} single-visit $r$-band 5$\sigma$ point source depth is expected to equal 24.6 mag, which is a significant increase over current surveys such as \gls{ATLAS} ($\sim19$ mag in $c$ and $o$ bands; \citealt{tonry_2018}) and \gls{ZTF} ($r \approx 20.6 - 20.9$; \citealt{bellm_2019}). This increased depth means that we will be able to probe faint precursor emission over a much larger volume. \citet{gagliano_2025} estimates that we will be able to find $\sim$40-130 precursors per year for SN IIP/IIL, and $\sim$110 precursors per year for SN IIn. They also find that the median detection distance for 2020tlf-like events with Rubin \gls{LSST} is $\sim74$ Mpc, while it is $\sim341$ Mpc for their more luminous IIn precursor models. Although they note that the exact numbers depend on the assumed model, it is clear that Rubin \gls{LSST} will significantly increase the sample size of SNe with detected precursor emission. Confidently quantifying the presence or absence of pre-SN variability in these light curves will be crucial. In addition, accurately quantifying detection thresholds and upper limits is essential, even when no variability is observed, as they can help constrain the physical mechanisms behind the presence of the \gls{CSM}.  



However, many SN for which precursor emission may be present are embedded in bright nearby galaxies. These bright host galaxies can introduce artifacts even in state-of-the-art image subtraction pipelines, which makes correctly identifying and quantifying pre-SN variability challenging. To this end, we developed \gls{DETECT}, which is able to characterize detection thresholds and is tailored to Rubin \gls{LSST}. The pipeline works by searching for locations in the host galaxy that are representative of the location in the galaxy where the SN appeared, and subsequently performing a series of synthetic source injection, image subtraction, and forced photometry. A similar pipeline has successfully been used before in the \gls{YSE} \citep{jones_2021} to study SN 2020tlf \citep{jacobson_galan_2022,jacobson_galan_2025} and SN 2023ixf \citep{ransome_2024}. 

It is worth noting that other methods to analyze light curves and look for precursor emission exist. For example, \gls{ATClean} has been used in the past to look for pre-SN variability in data from the \gls{ATLAS} survey \citep{tonry_2018} for SN 2023ixf \citep{rest_2025} and SN 2024ggi (G{\'e}ron et al. in prep.). However, both methods work very differently. \gls{ATClean} works by first cleaning the pre-SN light curve, and afterwards effectively convolving it with Gaussians of different kernel sizes to look for faint signals. Instead, \gls{DETECT} is built specifically to deal with sources that are embedded in bright host galaxies and works by carefully injecting fake sources to characterize the detection threshold.

Development for \gls{DETECT} started with simulated data from Rubin \gls{DP0} \citep{lsst_desc_2021}, and finished shortly after the release of Rubin \gls{DP1} \citep{rubin_dp1_2025}. This paper describes the different steps of the \gls{DETECT} pipeline and the results from applying it to data from Rubin \gls{DP1}. Specifically, in Section \ref{sec:data}, we describe the data used in this work. Then, we elaborate on the problem that \gls{DETECT} is built to address and describe the workflow of the \gls{DETECT} pipeline in greater detail in Section \ref{sec:pipeline_description}. In Section \ref{sec:results_dp0}, we describe the results from a set of tests of the \gls{DETECT} pipeline performed on simulated data from Rubin \gls{DP0}. We then apply the \gls{DETECT} pipeline to a sample of SN for which pre-SN data was available in Rubin \gls{DP1}. We discuss these results, summarize common challenges, and compare upper limits obtained by \gls{DETECT} to those published by Rubin in Section \ref{sec:results}. Finally, we summarize the results of this work in Section \ref{sec:conclusion}.

The results described here will aid in preparation for the start of the main \gls{LSST} survey. Finally, even though \gls{DETECT} was developed in the context of pre-SN variability, it can be used in any scenario where detections are uncertain and reliable upper limits are desired. We assume a flat $\Lambda$CDM cosmological model with H$_{0}$ = 67.7 km s$^{-1}$ Mpc$^{-1}$ and $\Omega_{\rm m}$ = 0.310 where necessary, based on \citet{planck_2020} and implemented with \texttt{Astropy} \citep{astropy_2013,astropy_2018,astropy_2022}.


\section{Data}
\label{sec:data}

\subsection{Rubin Data Preview 0}
\label{sec:data_rubin_dp0}

Development of \gls{DETECT} started with Rubin \gls{DP0}. This was the first publicly released data product by the Rubin Observatory. It is a simulated dataset designed to mimic the \gls{LSST} data products. \gls{DP0} is based on the \gls{LSST} \gls{DESC} \gls{DC2} simulated sky survey \citep{lsst_desc_2021}. The \gls{DESC} \gls{DC2} simulation produced \gls{LSST}-like images corresponding to five years of the planned ten year \gls{LSST} survey. It covered $\sim$300 deg$^{2}$ in the same six broad bands that will be used for the main survey, $ugrizy$, with an \gls{LSST}-like cadence. \gls{DP0} was then created by running the LSST science pipelines \citep{rubin_pipelines_2025} on these simulated images. It was meant as an early test for the \gls{LSST} science pipelines and to facilitate early development of external pipelines, such as \gls{DETECT}. While Rubin \gls{DP0} contains SNe that were injected as part of the \gls{DESC} \gls{DC2} simulation, in this work, we used the \gls{DP0} images themselves and injected our own galaxies and SNe to analyze in greater detail.

\subsection{Rubin Data Preview 1}
\label{sec:data_rubin_dp1}

\gls{DETECT} was subsequently tested on real data from Rubin \gls{DP1}. Specifically, we apply \gls{DETECT} at the location of several transients for which data was available in \gls{DP1} before their discovery date.

\gls{DP1} was the first set of real data released by the Rubin Observatory. Images were taken over an $\sim$1.5 month period between 24 October 2024 and 11 December 2024 \citep{rubin_dp1_2025, rubin_dp1_dataset}. The data in \gls{DP1} was taken using the \gls{LSSTComCam}, a smaller version of the actual \gls{LSSTCam} \citep{rubin_comcam_2024}, mounted on the Simonyi Survey Telescope in Cerro Pach\'on. The primary and tertiary mirrors of the telescope (M1M3) form a continuous surface with a diameter of 8.4m, while the secondary mirror (M2) has a diameter of 3.42m \citep{ivezic_2019}.

Rubin \gls{DP1} consists of observations of $\sim15$ deg$^{2}$ in seven fields: 47 Tucanae Globular Cluster, Extended Chandra Deep Field South, Euclid Deep Field South, Fornax Dwarf Spheroidal Galaxy, Low Galactic Latitude Field, Low Ecliptic Latitude Field, and the Seagull Nebula. These observations were made in the same six broad bands that the main \gls{LSST} will use \citep{ivezic_2019}. The cadence in \gls{DP1} varied strongly on the observed field: the Fornax Dwarf Spheroidal Galaxy field only has 2 epochs, while the Extended Chandra Deep Field South has 21 epochs. The 5$\sigma$ point source depth for the $r$-band coadded images for the different fields in \gls{DP1} varied between 24.19-25.96 mag.

The \gls{DP1} data was automatically reduced and analysed by the Rubin Science pipelines \citep{rubin_pipelines_2025}, which performed tasks including (but not limited to) coaddition of images, difference image analysis, and source detection. We accessed the final data products using the \gls{RSP} \citep{rubin_science_platform_2017, rubin_science_platform_2024}. Throughout this work, we regularly used the visit, coadded template, and difference images from Rubin \gls{DP1} \citep{rubin_dp1_visit, rubin_dp1_template, rubin_dp1_difference}.


The figures in this paper were created using data from Rubin \gls{DP1}, unless stated otherwise.

\subsection{ATLAS}
\label{sec:data_atlas}

While \gls{DETECT} is designed to work with data from Rubin \gls{LSST}, we also use data from the the \gls{ATLAS} survey \citep{tonry_2018} in this work to provide additional context for some of the SN targets whose \gls{DP1} data we analyze. While originally built to find \gls{NEA}, \gls{ATLAS} is often used in time domain astrophysics due to its high cadence ($\sim$2 days) and all sky coverage. \gls{ATLAS} operates four 0.5-m telescopes. Two are located in the southern hemisphere, at El Sauce in Chile and Sutherland in South Africa, while the other two are in Hawai`i, on Haleakala and Mauna Loa. \gls{ATLAS} obtains observations in the cyan ($c$) and orange ($o$) filters. The 5$\sigma$ point source depth for both bands is $\sim19$ mag \citep{tonry_2018}. The \gls{ATLAS} data presented in this work was obtained using the forced photometry server \citep{shingles_2021}.

In order to remove spurious epochs from the \gls{ATLAS} light curve, we remove all measurements where the measurement uncertainty ($\sigma_{f}$) exceeds \todo{160} $\mu$Jy or where the reduced chi-square of the PSF fits to the source apertures in the difference imaging ($\chi^{2}_{\rm PSF}$) is smaller than 10. Both of these thresholds for selecting high quality data were previously determined by \citet{rest_2025}. Finally, as \gls{ATLAS} typically obtains multiple images in a given band per night, we use the median value for each band for each day.


\section{Description of the \gls{DETECT} pipeline}
\label{sec:pipeline_description}

\subsection{Illustrating the problem}
\label{sec:illustration_problem}


New transient objects are often discovered using an image subtraction pipeline, where a template image is subtracted from a science image to produce a difference image. This difference image is then analyzed to identify transient or variable sources. In the default Rubin pipeline, a new transient object is flagged as a detection when the \gls{SNR} is greater than $5$ in the difference image at any given location. Image subtraction is a commonly used and powerful tool to quickly discover new objects in large amounts of data. However, it is also a complex problem that becomes increasingly more challenging when the target is embedded in a bright galaxy. This can introduce artifacts even in state-of-the-art image subtraction pipelines, such as the Rubin Science Pipelines. These artifacts can lead to false-positive detections, as well as generally impacting the sensitivity and depth at a given location within a Rubin image.

This problem can be visualized by looking at the Rubin \gls{DP1} data products for SN 2025brs. This target was reported by \citet{rehemtulla_2025} on 16 February 2025 at $\alpha = 06^{h}18^{m}59.11^{s}$, $\delta = -24^{\circ}37'38.93"$ based on images obtained by the \gls{ZTF}, and its pre-SN light curve was covered by Rubin \gls{DP1}. It is classified as a \gls{SN Type Ia} located in a nearby galaxy with a redshift of 0.00952 \citep{auchettl_2025}. The $r$-band Kron magnitude of the host galaxy is \todo{13.57} mag. 

\begin{figure*}
	\includegraphics[width=\textwidth]{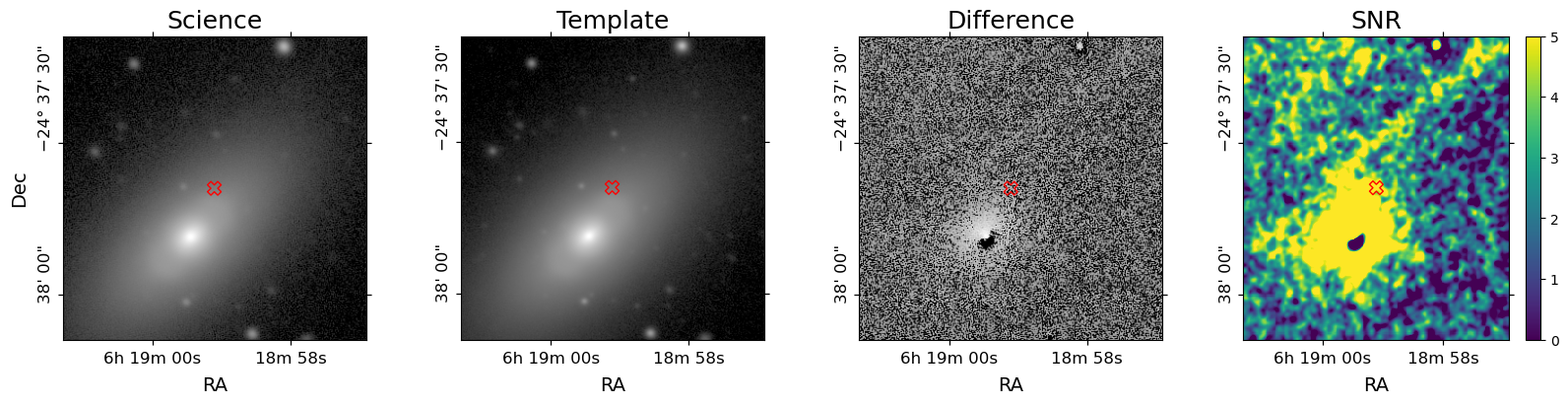}
    \caption{The $r$-band science image for SN \todo{2025brs} at MJD = \todo{60647} (left), as well as the corresponding $r$-band template (middle left) and difference images (middle right), taken from Rubin \gls{DP1}. The rightmost panel shows the \gls{SNR} calculated by performing forced photometry  on the difference image at every location. The red cross shows the reported location of SN \todo{2025brs}. It is clear that the image subtraction pipelines did not work as expected, presumably due to the bright galaxy that SN \todo{2025brs} is embedded in.}
    \label{fig:example_bad_subtraction}
\end{figure*}

For this illustration, we specifically focus on the DP1 image obtained at MDJ=60647 (3 December 2024; approximately 75 days prior to the discovery of the SN by \gls{ZTF}). The Rubin \gls{DP1} $r$-band science, template and difference images for this epoch are shown in Figure \ref{fig:example_bad_subtraction}. We also calculate the \gls{SNR} by performing forced photometry at every position in the difference image, which is shown in the rightmost panel of Figure \ref{fig:example_bad_subtraction}. The flux in the Rubin \gls{DP1} $r$-band difference image at this epoch equals \todo{$1650 \pm 292$ nJy} (or \todo{$23.4 \pm 0.2$ mag}). This corresponds to a \gls{SNR} of 5.65, which suggests that this would be a detection using the default \gls{SNR} $>5$ threshold. However, the rightmost panel of Figure \ref{fig:example_bad_subtraction} clearly shows that most of the host galaxy has a \gls{SNR} $> 5$. Therefore, the location of SN 2025brs, and most of the entire galaxy, will be considered a detection by the default pipeline, even though this is not necessarily correct.\footnote{Note that Rubin \gls{DP1} did not actually flag a detection (or ``diaSource'') at the exact coordinates of this target. However, 5 sources were identified within $5$ arcsec, and 24 sources within $10$ arcsec, over multiple epochs. This is likely due to the large area with \gls{SNR}$>5$ for this target (see the rightmost panel of Figure \ref{fig:example_bad_subtraction}), which complicates source detection. The flux values quoted here are obtained by performing forced photometry ourselves directly at the location of SN 2025brs.}

This type of issue is not limited to SN 2025brs, but it can be a problem for any target that is embedded in a bright galaxy. To visualize this point, we perform a test with the Rubin \gls{DP0} simulated images. We first inject a \gls{SDSS} \citep{york_2000, kollmeier_2019} $r$-band postage stamp of NGC 6308 into a part of the \gls{DP0} simulated Rubin sky. We then additionally inject a fake star in the outskirts of the injected galaxy. Finally, we used the Rubin pipelines to perform image subtraction and forced photometry on the difference image at the location of the injected star to measure the resultant \gls{SNR}. This process was then repeated multiple times, adopting a range of magnitudes for both the background galaxy and the injected star. 

\begin{figure}
	\includegraphics[width=\columnwidth]{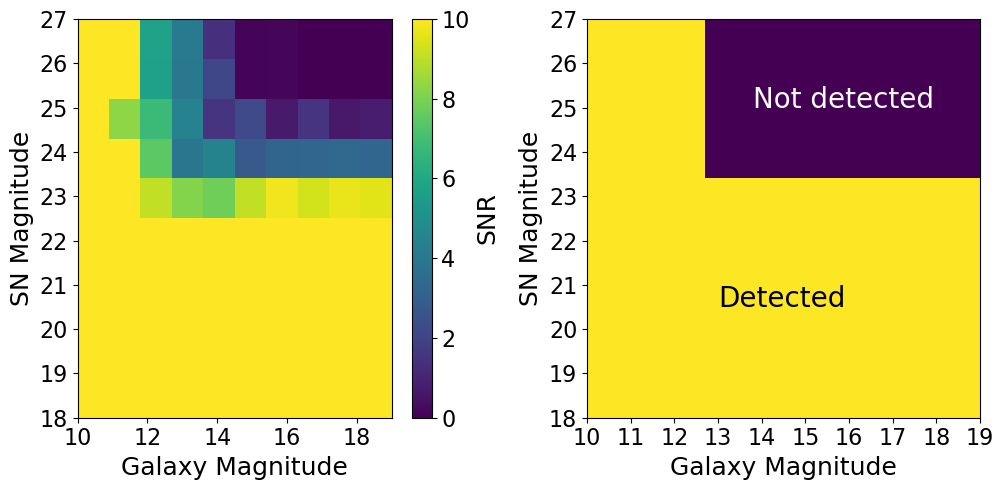}
    \caption{The left panel shows the \gls{SNR} of the difference image after injecting a galaxy and star in the science image, and only a galaxy in the template image of Rubin \gls{DP0}. This was done repeatedly for different magnitudes of the star and galaxy. The right panel shows which areas of this parameter space would be counted as detections using the default \gls{SNR} $> 5$ threshold. This shows that very faint targets that could not have been detected are incorrectly classified as detections when their host galaxies are very bright  ($\textrm{m}_{\textrm{gal}} \lesssim 13 \; \textrm{mag}$).}
    \label{fig:2d_naive}
\end{figure}

The results of this are shown in the left panel of Figure \ref{fig:2d_naive}, while the right panel shows which areas of the parameter space would be considered detections using the default \gls{SNR} $> 5$ threshold. When the galaxy is faint ($\textrm{m}_{\textrm{gal}} \gtrsim 15 \; \textrm{mag}$), the \gls{SNR} behaves as expected: low when the injected star is faint, and high when the injected star is bright. The transition region is around $\textrm{m}_{\textrm{star}} \approx 23-24 \; \textrm{mag}$, just below the expected 5$\sigma$ point source depth for the $r$-band for single exposures of 24.7 mag \citep{bianco_2022}. However, this overall behavior changes when the host galaxy becomes very bright ($\textrm{m}_{\textrm{gal}} \lesssim 13 \; \textrm{mag}$). Around this threshold, the measured \gls{SNR} at the location of the injected star is always larger than 5, regardless of the magnitude of the injected source. This is the case even for extremely faint injected stars ($\textrm{m}_{\textrm{star}} \sim 27 \; \textrm{mag}$) that clearly could not have been detected by Rubin in these single exposures. This means that the default Rubin pipelines would incorrectly flag these faint injected stars as detections. 

The results of this test indicates that we need to be more careful with characterizing transients that are embedded in bright nearby galaxies, since the default image subtraction pipelines can lead to misclassifications in these specific cases. In particular, while bright galaxies are most common at smaller distances (d $\lessapprox$ 100 Mpc), this is the same distance range over which Rubin is expected to be sensitive enough to find SN precursor emission (e.g., see \citealt{gagliano_2025}). Thus, additional caution is warranted when looking for precursor emission in nearby galaxies.

\subsection{Overview of the \gls{DETECT} workflow}
\label{sec:workflow}

As noted above, image subtraction is a complex problem, especially when dealing with bright nearby galaxies. We developed \gls{DETECT} to help tackle these issues. However, we emphasize that the pipeline is not an alternative for image subtraction. Instead, the goal of \gls{DETECT} is to quantify the detection efficiency and calculate accurate detection thresholds on a case-by-case basis (i.e., at a particular location within a specific Rubin image), even in situations where the image subtraction pipeline did not work as well as intended. In summary, the different steps of the \gls{DETECT} pipeline are:

\begin{enumerate}
    \item Identify locations within the host galaxy with similar background flux as the location of the target of interest. These locations are called ``injection locations''.
    \item Inject fake point sources of known magnitudes into these previously identified injection locations.
    \item Use the Rubin image subtraction pipelines to create a difference image.
    \item Perform forced photometry on the difference image and determine whether these fake sources are detected.
    \item Repeat steps 2 - 4 for different injection locations and a wide range of test source magnitudes.
    \item Approximate the background contamination at the location of the injected sources and target of interest.
    \item Finally, compile statistics on the fraction of injected sources that are recovered as a function of magnitude (which we refer to as the `recovery curve'), and fit a model to this data. This model can be used to obtain new detection thresholds tailored to the target of interest that can then be compared to the reported magnitude of the target to determine if we trust the detection.
\end{enumerate}


These steps are explained in greater detail in Sections \ref{sec:finding_injection_locations} - \ref{sec:fitting_recovery_curve}. We use $r$-band of SN \todo{2025brs} at MJD = 60647 as an example, since it is a nearby \gls{SN Type Ia} embedded in an extended and bright host galaxy where the image subtraction pipelines performed suboptimally, which makes it the ideal target for \gls{DETECT}. The code used in this work has been made publicly available on GitHub \href{https://github.com/tobiasgeron/detect}{here}.\footnote{https://github.com/tobiasgeron/detect} 

\subsection{Finding injection locations}
\label{sec:finding_injection_locations}

The first step of this process involves finding a set of locations, which we call injection locations, within the Rubin image that can be used to characterize the detection threshold at the location of a target of interest. Specifically, we select locations that are both (i) nearby (by default \todo{$<50$} arcsec, but this can be changed by the user) and (ii) have similar flux from the host galaxy as the target in question. The implicit assumption is that sites within a galaxy with similar background flux will be affected in a similar way by the default Rubin image subtraction pipelines. The injection locations are found by first smoothing the template exposure, and then finding locations that have flux values similar to the site of the target in the smoothened template exposure. By default, the smoothing is done by calculating the median in a sliding window with a size of \todo{10} pixels, while the background flux at the injection locations is required to be within \todo{5}\% of the background flux at the location of the target of interest. However, both of these parameters can be changed by the user. \gls{DETECT} will identify 10 injection locations by default, but this can also be changed by the user with the \texttt{n\_injections} parameter. Please see Appendix \ref{app:different_settings} for more details on how to select the right value for your use case. 

Since the upcoming image subtraction step (see Section \ref{sec:injection_subtraction_photometry}) is the most time-intensive step in the \gls{DETECT} pipeline, we attempt to save time by injecting multiple sources simultaneously (in a so-called ``injection iteration''). However, this means that we need to be careful with selecting injection locations. We want to avoid excessive overlap between the \glspl{PSF} of different injected sources, as this could bias the subsequent results (i.e., if enough light from one injected source ``leaks'' into the region covered by another it can influence whether a source is recovered in subsequent steps). This is done by looking at the \gls{PSF} associated with each possible injection location and calculating the distance at which 99.9\% of the flux of the normalized PSF is contained.\footnote{The \gls{PSF} can vary across the field of view of typical Rubin images. One of the data products provided by the Rubin pipelines is a model for the PSF at each location within a given image, which we use in \gls{DETECT}.} Injection locations are combined only if they are separated by this distance from each other and from the original target in question. Other methods to avoid leaking between injected sources are also incorporated into \gls{DETECT}. For example, it can calculate the $\sigma$ of the \gls{PSF} and only select injection locations that are separated by a multiple of that value (which the user can specify). The user also has the option to disable the simultaneous injection of sources entirely, at the cost of compute time.

\begin{figure}
	\includegraphics[width=\columnwidth]{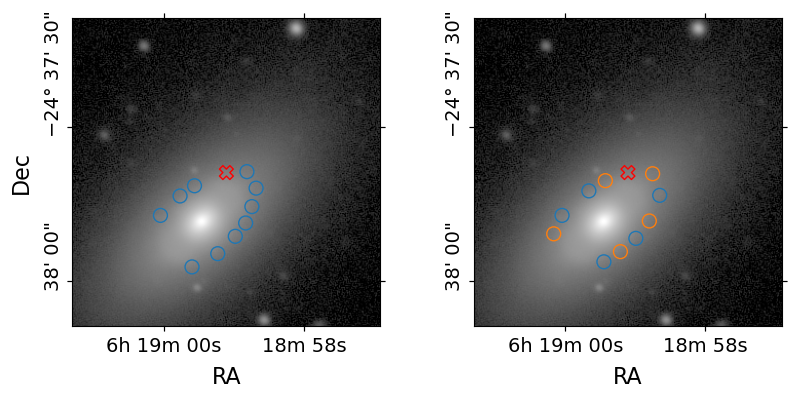}
    \caption{The left panel shows possible injection locations for SN \todo{2025brs} with the default \gls{DETECT} settings. Note that in this case the injection locations naturally fall on an elliptical isophotal contour; however DETECT is written to identify appropriate injection locations for arbitrarily complex galaxy morphologies. Since the host galaxy is large and nearby, \gls{DETECT} was able to find locations that are sufficiently spread out so they can all be injection together (i.e. they are in the same injection iteration). The injection locations in the right panel were selected with more strict criteria, and are therefore more spread out and divided over two different injection iterations, indicated by their color. The red cross in both panels indicates the location of SN \todo{2025brs}.}
    \label{fig:injection_locs}
\end{figure}

The injection locations found for SN \todo{2025brs} are shown in Figure \ref{fig:injection_locs}. Since we select sites with similar flux values, these injection locations will typically be located on isophotal ellipses for galaxies with relatively smooth surface brightness distributions. However, the distribution of injection locations will be more diverse for galaxies with complex morphology. The left panel shows a set of 10 injection locations that all meet the default criteria of being separated by more than the distance that contains 99.9\% of the flux in the normalized PSF at each location. As a result, sources will be injected in all locations as part of the same injection iteration. In contrast, the injection locations in the right panel were selected using stricter criteria (i.e., injection locations in the same injection iteration need to be separated by more than the distance that contains 99.99\% of the flux in the normalized PSF as well as more than $3\sigma$ of the \gls{PSF} at each location). As a result, \gls{DETECT} separates the 10 injection locations into two distinct injection iterations, visualized by color.

Finding injection locations is relatively straightforward for large nearby galaxies, such as the host galaxy of SN \todo{2025brs}. Luckily, this is also the type of situation where DETECT is most likely to be used, as it was specifically designed to handle targets embedded in bright nearby hosts. However, it is still possible to find appropriate injection location for smaller galaxies, although in general more injection iterations and some finetuning of the settings are required. Please refer to Appendix \ref{app:suboptimal_inj_locs} for some examples.

\subsection{Source injection, image subtraction, and forced photometry}
\label{sec:injection_subtraction_photometry}

\begin{figure}
	\includegraphics[width=\columnwidth]{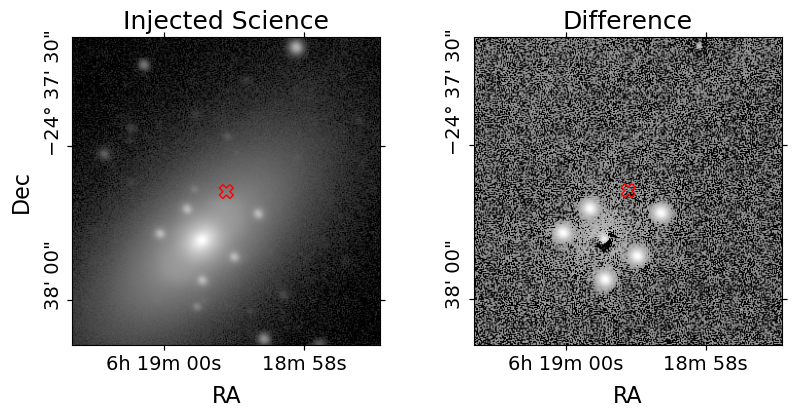}
    \caption{The left panel shows the science image of SN \todo{2025brs} with bright ($m_{r} = 18$ mag) sources injected in the locations of one of the injection iterations shown in the right panel of Figure \ref{fig:injection_locs}. The right panel shows the difference image after performing image subtraction. The red cross in both panels indicates the location of SN \todo{2025brs}.}
    \label{fig:inj_image_subtraction}
\end{figure}

The next step involves using the Rubin source injection pipelines to inject point sources at the selected injection locations and the Rubin image subtraction pipelines to create the resultant difference image. Note that, while the injection locations were selected based on the template image, we inject the point sources into the individual science images. We use Rubin's \texttt{VisitInjectTask} for the source injection and \texttt{AlardLuptonSubtractTask} for the image subtraction.\footnote{Unlike in \gls{DP0}, the pipeline configuration and information needed to exactly reproduce the \gls{DP1} difference images was not shared with the wider astronomical community. However, once this is shared, presumably with \gls{DP2} or \gls{DR1}, we will update the published version of \gls{DETECT} to reflect this.} An example of this process is shown in Figure \ref{fig:inj_image_subtraction}, where the left panel shows a science image with a set of test sources injected in the injection locations calculated above (i.e., the blue injection locations from right panel of Figure \ref{fig:injection_locs}). The right panel of Figure \ref{fig:inj_image_subtraction} shows the resultant difference image after performing image subtraction. For this demonstration, we inject bright sources of $m_{r} = 18$ mag to make visualization clearer. As this is well above the nominal Rubin detection threshold, they are clearly visible in the difference image. However, we emphasize that this analysis would usually be performed for fainter sources. Consequently, in many cases where \gls{DETECT} is used, the sources will not necessarily be as clearly visible as they are here. We then perform forced photometry (using Rubin's \texttt{ForcedMeasurementTask}) on these locations in the difference image and assess whether the injected sources are detected using the default \gls{SNR} $>5$ threshold (pending a background correction, see Section \ref{sec:background}).

The \gls{DETECT} pipeline then repeats the process described above for a wide range of injected source magnitudes and calculates what fraction of injected sources were recovered at each magnitude tested. We expect all sources to be detected when bright sources are injected, and no sources to be detected when very faint sources are injected, with a transition region in between. It is crucial to sample this transition region to obtain a  recovery curve and detection thresholds that are reliable (see Section \ref{sec:fitting_recovery_curve}). \gls{DETECT} samples the magnitude space by first testing a range of magnitudes around the initial reported magnitude of the target in question, and then follows an iterative process to more densely sample the magnitude space around the transition region. This process can also be finetuned by the user with the \texttt{n\_mag\_steps} parameter; please see Appendix \ref{app:different_settings} for more detail.

\subsection{Characterizing and Correcting for Background Contamination}
\label{sec:background}

There is one remaining issue that must be taken into account in order to obtain an accurate recovery curve: potential background contamination. To illustrate this point, we refer again to Figure \ref{fig:example_bad_subtraction}, which showed an epoch where the Rubin image subtraction pipeline did not work perfectly for SN 2025brs. This implies that some of the flux in the difference image is coming from the host galaxy. If this is not corrected for, all injected sources, regardless of their magnitude, will always be counted as detections (indeed, this is closely related to the problem that \gls{DETECT} was designed to address, as described in Section \ref{sec:illustration_problem}).

For the injected sources, \gls{DETECT} attempts to correct for this when determining whether a given injected source is actually detectable via a two step process. First, in order to estimate the level of background contamination, forced photometry is performed at the injections locations in the difference images prior to injecting any fake sources. The resulting flux is then subtracted from the flux that is measured via forced photometry on the difference image after the source is injected, while correctly propagating uncertainties. This (lower) recovered flux is then used to determine whether a given injected source was detected at $>5\sigma$. The implicit assumption is that the level of background contamination in the difference image is largely unaffected by the presence of the injected sources. This is done automatically by \gls{DETECT}, although the user is able to disable this behavior. In addition, flags are returned by \gls{DETECT} to indicate in whether significant background contamination was identified.

This issue of background contamination applies to the initial reported magnitude of the transient of interest as well. However, we cannot apply the same method as described above to correct for this, as we cannot directly disentangle what fraction of the flux in the original difference image is due to the target or due to artifacts from the host galaxy. However, \gls{DETECT} does provide some tools to approximate this. In particular, \gls{DETECT} will identify all locations in the template that have similar flux as the location of the target of interest, while also being separated from the target by more than the distance that contains 99.9\% of the flux of the normalized \gls{PSF}.\footnote{Note that this typically includes the previously identified injection locations, though we also include other locations with similar flux levels. This is done so that we can average over more locations and be more accurate in our estimate of the background contamination.} \gls{DETECT} then performs forced photometry on the difference image at \emph{all} of these locations to approximate the flux in the difference image due to artifacts from the host galaxy. \gls{DETECT} will then calculate the median from this distribution and subtract that flux from the reported flux for the target of interest. The user can adjust which statistic is used (e.g., mean or median) when calculating the background flux based on the background flux distribution found using the set of test locations. When making these corrections, we assume that areas with similar host galaxy flux will be similarly affected by the image subtraction pipelines and that there are no other transient sources in the set of test locations.

\begin{figure}
	\includegraphics[width=\columnwidth]{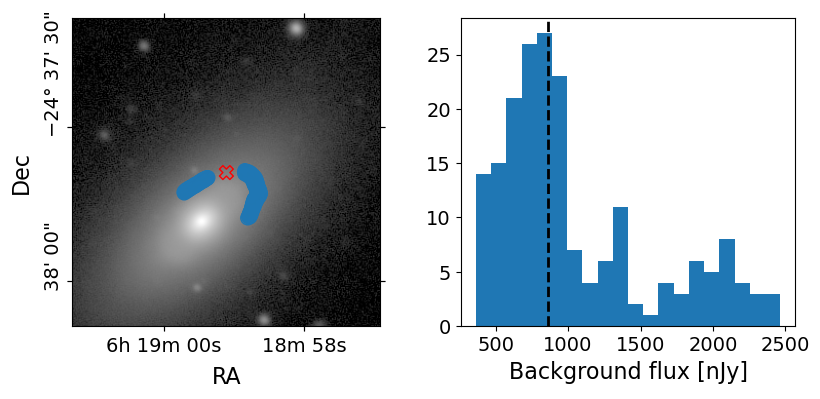}
    \caption{The blue region in the left panel shows the locations in the host galaxy where the flux in the template is within 5\% of the flux in the template at the SN location. The red cross shows the location of SN \todo{2025brs}. The right panel shows the distribution of fluxes that are found by performing forced photometry on the difference image at all of these locations. The dashed vertical line indicates the median of this distribution, which is equal to \todo{$863\pm307$} nJy.}
    \label{fig:background}
\end{figure}

We demonstrate this process applied to the MJD = 60647 epoch of SN \todo{2025brs} in Figure \ref{fig:background}. The left panel shows the region in the galaxy where the flux in the template is within 5\% of the flux in the template at the SN location. The right panel shows the distribution of fluxes found by performing forced photometry at all these locations in the difference image. This distribution does not always have a Gaussian shape. Thus, the user should be careful with which statistic to use, and it is recommended to approach this on a case-by-case basis. For this example, we proceed with the median of the distribution, which is equal to \todo{$863\pm307$} nJy. Subtracting that from the initial measurement and converting to magnitudes changes the initial value of SN \todo{2025brs} from \todo{$23.4\pm0.2$} mag to \todo{$24.2\pm0.58$} mag.

In the Sections below where we apply the \gls{DETECT} pipeline to both \gls{DP0} (Section \ref{sec:results_dp0}) and \gls{DP1} (Section \ref{sec:results}) we use a strategy to mitigate possible effects of background contamination similar to that described above for SN \todo{2025brs}. Specifically, we compute the median of the distribution and if the median is equal to or lower than 0, no flux correction is applied. If it is larger 0, we correct the flux by subtracting this value from the original flux measurement.

\subsection{Recovery curve modeling}
\label{sec:fitting_recovery_curve}


After following the steps outlined in Sections \ref{sec:finding_injection_locations}, \ref{sec:injection_subtraction_photometry} and \ref{sec:background}, \gls{DETECT} provides us with the fraction of injected sources that are recovered at $>5\sigma$ over a wide range of magnitudes, sampled most closely around the transition region. However, we want to obtain actual magnitude detection threshold values. The current version of \gls{DETECT} does this by fitting a simple piecewise linear model to the data.\footnote{By default, \gls{DETECT} fits the recovery curve using the piecewise linear model described here. However, it is also possible to fit any other model the user thinks is appropriate, such as a Gauss error function, with the output provided by \gls{DETECT}.} First, we define two important transition points: $x_1$ and $x_2$. At magnitudes brighter than $x_1$, the recovery curve equals 1, i.e. the injected signal is always detected. At magnitudes fainter than $x_2$, the recovery curve equals 0, i.e. the injected signal is never detected. The area between $x_1$ and $x_2$ is fit with a line. Our model is summarized by: 

\begin{align}
    y &= 
    \begin{cases}
        1, & \text{if } x < x_1, \\
        mx + b, & \text{if } x_1 \leq x \leq x_2, \\
        0, & \text{if } x > x_2,
    \end{cases}
    \label{eq:piecewise}
\end{align} 

where $x$ is the magnitude of the injected source and $y$ is the detection probability. Furthermore:

\begin{align}
    m &= \frac{-1}{x_2 - x_1}, \quad
    b = \frac{x_2}{x_2 - x_1}.
    \label{eq:piecewise_2}
\end{align} 

\begin{figure}
	\includegraphics[width=\columnwidth]{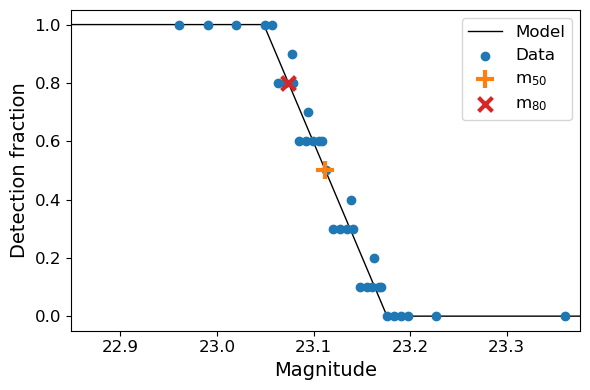}
    \caption{The empirical recovery curve (blue dots) as well as the fitted model (black line) are shown in this figure. We also calculate the 50\% and 80\% detection thresholds (m$_{50}$ and m$_{80}$), indicated with the orange upright cross (`+') and red diagonal cross (`$\times$'). }
    \label{fig:recovery_curve}
\end{figure}

This has the additional advantage that we can solve for any detection threshold, even if we did not explicitly sample it. \gls{DETECT} will solve for the 50\% and 80\% detection thresholds by default, but this can be changed by the user. For the case of the MJD = 60647 epoch of SN \todo{2025brs}, the 50\% and 80\% detection thresholds are \todo{23.11} and \todo{23.07}, respectively. As the corrected magnitude (and even the uncorrected magnitude) of SN \todo{2025brs} is fainter than both these thresholds, we can confidently conclude that no significant detection was made at this location in this epoch. The empirical recovery curve and the model for this example are shown in Figure \ref{fig:recovery_curve}.

\section{Initial Tests of the Pipeline on Simulated Rubin DP0 Data}
\label{sec:results_dp0}

After development, we performed a number of tests on the \gls{DETECT} pipeline using simulated data from \gls{DP0}. We first ran \gls{DETECT} for a wide range of host galaxy and target magnitudes (Section \ref{sec:results_dp0_2dplot}), both to test the pipeline and assess detection thresholds as a function of host galaxy brightness. We then inject a realistic pre-SN light curve into simulated DP0 images (Section \ref{sec:results_dp0_2009ip}) in order to assess the recoverability of these features even at large distances (e.g., D$>$100 Mpc).

\subsection{Detection Thresholds from \gls{DETECT} Across Host Galaxy and Transient Magnitudes}
\label{sec:results_dp0_2dplot}

\begin{figure}
	\includegraphics[width=\columnwidth]{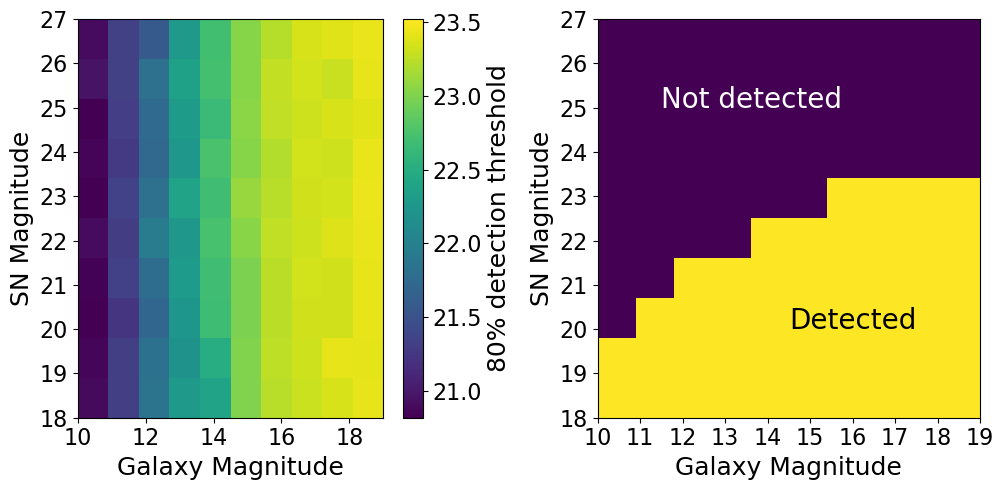}
    \caption{The left panel shows the 80\% detection threshold (i.e., the magnitude where 80\% of injected test sources were recovered at $\geq5\sigma$) calculated by \gls{DETECT} after injecting a galaxy and star in the science image, and only a galaxy in the template image in Rubin \gls{DP0}. This was done repeatedly for different magnitudes of the star and galaxy. The right panel shows which areas of this parameter space would be counted as detections by comparing the 80\% detection threshold to the magnitude of the injected star.}
    \label{fig:2d_detect}
\end{figure}

To test \gls{DETECT} for a wide range of host galaxy and target magnitudes, we inject fake sources and a \gls{SDSS} $r$-band postage stamp of \todo{NGC 6308} into a part of the \gls{DP0} simulated Rubin sky, similar to what was done in Section \ref{sec:illustration_problem} and Figure \ref{fig:2d_naive}. As before, we inject the source in the outskirts of the galaxy. However, instead of measuring the \gls{SNR} from performing forced photometry at the source location in the difference image, we ran \gls{DETECT} in order to calculate the corresponding 80\% detection thresholds. This was repeated for a range of magnitudes for the galaxy and injected source. The 80\% detection threshold for every combination is shown in the left panel of Figure \ref{fig:2d_detect}. Note that, while there is some variability in the vertical direction in this plane, most of the variability is in the horizontal direction. This means that the magnitude of the host galaxy affects the value of the 80\% detection threshold much more than the magnitude of the source. It is also worth noting that, across the parameter space tested here, the 80\% detection threshold varies by more than 2 magnitudes. It is equal to $\gtrsim23$ mag when the host galaxy is faint ($\textrm{m}_{\textrm{gal}} \gtrsim 16 \; \textrm{mag}$) and equal to $\sim21$ mag when the host galaxy is bright ($\textrm{m}_{\textrm{gal}} \lesssim 12 \; \textrm{mag}$). This suggests that it is easier to detect an event if its host galaxy is fainter and highlights again that it is important to be careful when characterizing transients that are embedded in bright galaxies.

The right panel of Figure \ref{fig:2d_detect} shows in which area of this parameter space the magnitude of the injected source is brighter than the 80\% detection threshold, and is therefore counted as a detection. In contrast to Figure \ref{fig:2d_naive}, this now behaves much more as expected. When the host galaxy is faint ($\textrm{m}_{\textrm{gal}} \gtrsim 16 \; \textrm{mag}$), the border between detections and non-detections is around $\textrm{m}_{\textrm{star}} \approx 23-24 \; \textrm{mag}$. However, when the host galaxy becomes brighter ($\textrm{m}_{\textrm{gal}} \lesssim 15 \; \textrm{mag}$), it becomes increasingly more difficult to detect an event, and the border between detections and non-detections is pushed to lower magnitudes (i.e. the event must be brighter before we can confidently detect it).  

We emphasize that the exact results shown in Figure \ref{fig:2d_detect} will vary depending on the morphology of the host galaxy, location of the transient source, as well as the quality of the templates used for image subtraction. This section is not meant to exhaustively characterize this full parameter space, but rather to provide an overview of the detection threshold trends with host galaxy and transient magnitude.

\subsection{Applying \gls{DETECT} to a Realistic Pre-SN Light Curve}
\label{sec:results_dp0_2009ip}



\begin{figure}
	\includegraphics[width=\columnwidth]{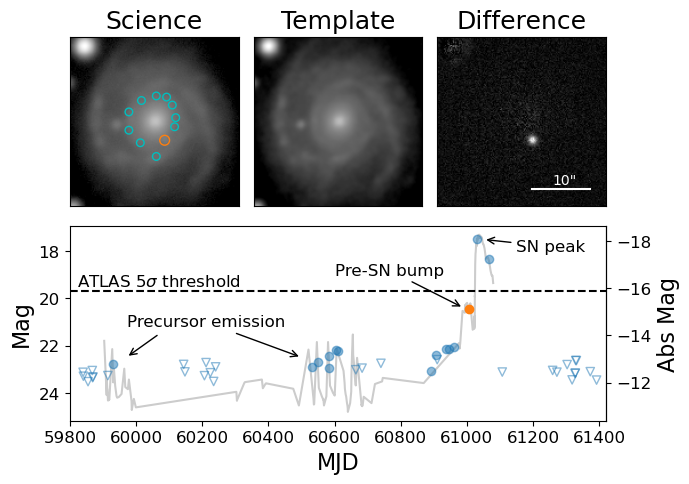}
    \caption{We injected a postage stamp of \todo{NGC 6308} and the light curve of SN 2009ip into Rubin \gls{DP0} at a distance of \todo{129.74} Mpc to simulate whether we can recover it at this distance using \gls{DETECT}. The top rows shows the science, template, and difference image of the epoch corresponding to the pre-SN bump phase of SN 2009ip. The science image additionally shows the location of where SN 2009ip was injected (orange circle) and the injection locations identified by \gls{DETECT} (cyan circles). The difference image shows a scale bar of 10 arcsec. The bottom row shows the rescaled full light curve of SN 2009ip (grey line), as well as the Rubin \gls{DP0} epochs in which it was injected. Full circles denote epochs where the reported flux exceeded the 80\% detection threshold (i.e., these epochs are detected), whereas downward triangles represent non-detections, placed at the 80\% detection threshold. The dashed horizontal line denotes the 5$\sigma$ detection threshold of \gls{ATLAS} at $19.7$ mag \citep{tonry_2018}.}
    \label{fig:2009ip}
\end{figure}

While we have shown in Section \ref{sec:results_dp0_2dplot} that the detection threshold can vary dramatically depending on the magnitude of the host galaxy, Rubin \gls{LSST} is still expected to be a very powerful probe of pre-SN variability and outbursts. This is demonstrated with SN 2009ip, a SN Type IIn with a well-established precursor outburst $\sim$40 days before explosion. It also showed additional precursor emission in 2009 during its \gls{LBV} phase \citep{pastorello_2013, prieto_2013, margutti_2014}. In Figure \ref{fig:2009ip}, we injected the light curve of SN 2009ip into Rubin \gls{DP0} images on top of a bright host galaxy to see whether we can recover its pre-SN variability at larger distances using \gls{DETECT}.

More specifically, we injected an \gls{SDSS} $r$-band postage stamp of NGC 6308 into a part of the simulated \gls{DP0} images. NGC 6308 has a distance of $129.74\pm9.08$ Mpc and $r$-band magnitude of 13.07 mag \citep{ablareti_2017}. We arbitrarily chose to inject it in $\alpha = 03^{h}46^{m}29.28^{s}$, $\delta = -36^{\circ}29'16.80"$. This part of the sky was simulated 131 times over the course of 1594 days in \gls{DP0}. After selecting a maximum of one epoch per day and removing epochs near the edges of their exposure, we are left with a total of 44 simulated $r$-band epochs. We then injected the $r$-band light curve of SN 2009ip $3$ arcsec offset from the center of its host galaxy in both right ascension and declination. The full light curve of SN 2009ip was obtained from \citet{pastorello_2013,prieto_2013,margutti_2014} and was shifted 4829 days towards the simulated \gls{DP0} date range. We interpolated the full light curve to determine the appropriate flux to inject at each \gls{DP0} epoch. The flux of SN 2009ip was rescaled from its original distance of 24 Mpc to a synthetic distance of 129.74 Mpc to match the distance of NGC 6308. Finally, these epochs were analyzed with \gls{DETECT}.

We see in the bottom row of Figure \ref{fig:2009ip} that the shifted and rescaled peak of SN 2009ip is clearly recovered, as it is $\sim5.7$ mag brighter than the 80\% detection threshold from \gls{DETECT}. We also detect the precursor outburst $\sim$40 days before explosion, which is $\sim2.3$ mag brighter than the 80\% detection threshold. We also find significant detections in a few earlier epochs with precursor emission during its \gls{LBV} phase. This precursor emission and outburst would have gone unnoticed by \gls{ATLAS}, which has a 5$\sigma$ detection threshold of 19.7 mag \citep{tonry_2018}.

This demonstrates that Rubin \gls{LSST}, together with \gls{DETECT}, is able to recover precursor emission for SN 2009ip-like events, even at $\sim 130$ Mpc. However, we emphasize that this is not meant as an exhaustive parameter search to quantify exactly how far Rubin \gls{LSST} will be able to observe these types of events in all possible conditions. Instead, we simply want to highlight the potential of Rubin \gls{LSST} and illustrate how \gls{DETECT} will be used.







\section{Application of \gls{DETECT} to search for precursor emission in Rubin DP1}
\label{sec:results}

In this Section, we test \gls{DETECT} on real data from Rubin \gls{DP1}, with the goals of (i) searching for precursor emission for a sample of transients that had their pre-discovery period covered by \gls{DP1} and (ii) identifying challenges that may arise when running \gls{DETECT} on real data. When doing so, we reiterate that the goal of \gls{DETECT} is not to perform image subtraction itself. Instead, we acknowledge that the current state-of-the art image subtraction pipelines are not perfect in certain situations and do a more detailed analysis to assess whether a detection is real. \gls{DETECT} does this by calculating detection thresholds tailored to a specific location within a host galaxy via a series of source injection, image subtraction, and forced photometry as described in Section \ref{sec:pipeline_description}.



In Section \ref{sec:sample}, we describe the sample of transients identified in the Rubin \gls{DP1} fields that we analyze with \gls{DETECT} in this work, while we summarize the settings of \gls{DETECT} that we used in Section \ref{sec:detect_dp1_settings}. The results are shown in Section \ref{sec:results_detections} for targets where we found evidence for a significant detection, and in Section \ref{sec:results_nondetections} for targets without any reliable detections. We outline common challenges that we faced during this analysis in Section \ref{sec:challenges} and make a comparison of the upper limits obtained with \gls{DETECT} to the default Rubin 5$\sigma$ upper limits in Section \ref{sec:comparison_upper_limits}.

Finally, it is worth noting that the upper limits and detection thresholds shown here were determined using the template coadded images available at the time of writing. The templates are expected to improve in quality over time after the main \gls{LSST} survey starts, which means that we will likely be able have more constraining upper limits in the future. 




\subsection{Sample Selection}
\label{sec:sample}

We tested our pipeline on data from Rubin \gls{DP1}. The targets were selected by first downloading all recorded entries on the \gls{TNS} on \todo{30 July 2025} (232 days after the end of DP1), resulting in a total of \todo{169,614} possible targets. A subset of this, \todo{159} targets, were actually located in the DP1 footprint. However, since we are primarily interested in pre-SN variability, we only kept targets that were discovered after the start of DP1 (24 October 2024), so that their precursor phase is possibly covered by DP1. This further reduced the sample to \todo{36} targets. 

We then ran \gls{DETECT} on a target if it met one of the following two criteria. First, we kept a target in our sample if the default Rubin pipelines flagged any detections at the location of the target (within \todo{1} arcsec). This was the case for \todo{21} targets. We also included targets for which a transient classification was available on \gls{TNS}, regardless of whether Rubin itself reported a source at that location. This was only the case for \todo{one} more target: SN 2025brs, the \gls{SN Type Ia} that was used as an example in Section \ref{sec:pipeline_description}. Note that, while the default Rubin pipelines did not flag any sources at the exact location of this target that were within 1 arcsec, they did identify 5 sources within 5 arcsec, and 24 sources within 10 arcsec. This is likely due to the large footprint of SN 2025brs with \gls{SNR} $> 5$ (see Figure \ref{fig:example_bad_subtraction}), which complicates source detection and localization. Including this target brings the sample size to \todo{22}. 


However, of these, \todo{7} targets were subsequently excluded from this work as they were located in faint galaxies that were relatively small, with $r$-band radii smaller than $\sim6$ arcsec. It was hard to find appropriate injection locations for these smaller galaxies, preventing subsequent analysis with \gls{DETECT}. In principle, we could find suboptimal injection locations for some of these targets by relaxing some of the pipeline parameters (e.g., see Section \ref{sec:finding_injection_locations}), but this is not advised, as the results from \gls{DETECT} will also not be ideal. However, we emphasize that these are also not the types of systems for which \gls{DETECT} was initially developed: it was designed to handle targets in bright nearby galaxies that may cause challenges for the default image subtraction pipelines. This reduced the final sample size to \todo{15} targets.

The final sample is summarized in Table \ref{tab:sample}. Four of these targets were discovered after the Rubin \gls{DP1} campaign had ended, while 11 were discovered during the campaign. Note that only two targets in this sample have confirmed redshifts: AT 2024aigg at $z=0.07593$ ($D_L = 355.5$ Mpc) and SN 2025brs at $z=0.00952$ ($D_L = 42.5$ Mpc). Additionally, only one target has a confirmed type: SN 2025brs is a \gls{SN Type Ia}. As a result, we highlight that some of these transients may not be SNe (although they are all are coincident with galaxies in the Rubin images, as noted above). In addition, full assessment of implications of possible pre-SN emission would require distances measurements. Nonetheless, this sample still provides a useful test case for the \gls{DETECT} pipeline on real Rubin data.


\begin{deluxetable*}{c c c c c c c c}
  \tablecaption{This table summarizes the \todo{15} targets analyzed in this work. We note its name in \gls{TNS}, RA, Dec, discovery date, the number of days between the start of \gls{DP1} and the discovery date ($t_{\textrm{discovery}} - t_{\textrm{start DP1}}$), an identifier that links back to Rubin \gls{DP1} (\texttt{diaObjectId}) , the reporting source, and reference. The table is sorted by discovery date.}
  \label{tab:sample}
  \tablehead{
  \colhead{TNS name} & \colhead{RA} & \colhead{Dec} & \colhead{Discovery date} & \colhead{$t_{\textrm{discovery}} - t_{\textrm{start DP1}}$} & \colhead{diaObjectId} & \colhead{Reporting Source} & \colhead{Reference}}
  \decimalcolnumbers
  \startdata
    AT 2024aaux & 38.594742 & 7.214531 & 2024-11-08 & 15 & 648374722634973207 & ZTF & \citet{sollerman_2024} \\
    AT 2024aigw & 52.731551 & -27.866353 & 2024-11-09 & 16 & 611255210081255575 & Rubin & \citet{freeburn_2025} \\
    AT 2024aigg & 53.124768 & -27.739815 & 2024-11-09 & 16 & 611255759837069401 & Rubin & \citet{freeburn_2025b} \\
    AT 2024aigh & 59.324176 & -48.368972 & 2024-11-18 & 25 & 592915218690998602 & Rubin & \citet{freeburn_2025c} \\
    AT 2024aigs & 59.221811 & -49.105017 & 2024-11-18 & 26 & 591819074317582360 & Rubin & \citet{freeburn_2025d} \\
    AT 2024ackk & 40.168017 & -34.297036 & 2024-11-27 & 34 & 604064060438217244 & ATLAS & \citet{tonry_2024b} \\
    AT 2024ahyy & 52.892587 & -28.412603 & 2024-12-02 & 39 & 609781520902651937 & YSE & \citet{murphey_2025} \\
    AT 2024ahzc & 52.838254 & -28.279900 & 2024-12-02 & 39 & 609782208097419314 & YSE & \citet{murphey_2025} \\
    AT 2024aigj & 53.212563 & -27.681277 & 2024-12-02 & 39 & 611256447031836769 & Rubin & \citet{freeburn_2025f} \\
    AT 2024aigi & 53.400537 & -28.543906 & 2024-12-04 & 41 & 609788117972418615 & Rubin & \citet{freeburn_2025g} \\
    AT 2024aigl & 59.850663 & -48.780701 & 2024-12-09 & 46 & 592913706862510093 & Rubin & \citet{anumarlapudi_2025} \\
    AT 2024ahsx & 53.366975 & -28.215100 & 2024-12-19 & 56 & 611253629533291838 & YSE & \citet{murphey_2025} \\
    AT 2024ahzi & 58.335054 & -48.750303 & 2024-12-19 & 56 & 592914119179370575 & YSE & \citet{murphey_2025} \\
    SN 2025brs & 94.746277 & -24.62748 & 2025-02-16 & 115 & - & ZTF & \citet{rehemtulla_2025} \\
    AT 2025sfa & 38.224495 & 7.261514 & 2025-07-24 & 273 & 648374860073926766 & GOTO & \citet{oneill_2025} \\
  \enddata
\end{deluxetable*}

\subsection{Overview of \gls{DETECT} settings used on \gls{DP1} data}
\label{sec:detect_dp1_settings}

We run the \gls{DETECT} pipeline on images from Rubin \gls{DP1} for the \todo{15} transients described in Section \ref{sec:sample}. We only analyze one epoch per day per band for computational reasons, and we always choose the highest \gls{SNR} epoch.\footnote{Rubin \gls{DP1} imaged the same part of the sky with the same band multiple times each night as part of its observing strategy. By only using the highest \gls{SNR} epoch per band per night, we reduce the required number of epochs we need to analyze by 50-90\%, depending on the target. For more information on the Rubin \gls{DP1} cadence, see \citet{rubin_dp1_2025}.} We need the photometry on the difference images to run our pipeline. For the \todo{14} objects in our sample where Rubin identified a source within \todo{1} arcsec in at least one epoch, this was obtained by querying the \texttt{ForcedSourceOnDiaObject} table for these objects \citep{rubin_dp1_forcedsourceondiaobject_table}. For SN 2025brs (for which Rubin did not identify a significant source within $1$ arcsec) we performed forced photometry on the difference image ourselves.

We used many of the default settings of \gls{DETECT}. Most importantly, we always find 10 injection locations (\texttt{n\_injection}) that are within 5\% of the background flux at the location of the target of interest (\texttt{p\_threshold}). We only inject fake sources simultaneously when they can be separated by the distance at which 99.9\% of the flux of the normalized \gls{PSF} is contained (\texttt{psf\_flux\_threshold}). We sample the transition region of the recovery curve carefully by iteratively zooming in (\texttt{n\_mag\_steps}). We speed up by the code by creating cutouts of 400$\times$400 pixels (\texttt{cutout\_size}) and parallelizing over the four CPU cores available on the \gls{RSP} (\texttt{n\_jobs}). See Section \ref{sec:finding_injection_locations} and \ref{sec:injection_subtraction_photometry} for a more detailed overview of these parameters. See Appendix \ref{app:different_settings} for advice on how to finetune these parameters for your use-case and see Appendix \ref{app:suboptimal_inj_locs} for advice on how to find injection locations for targets in embedded in smaller galaxies. We applied a background correction by identifying regions in the template with similar flux to the location of the target. We then perform forced photometry at those locations in the difference image and take the median. If the median is positive, we subtract it from the original target flux measurement (see Section \ref{sec:background}).
 
Finally, we calculate the 80\% detection threshold for the identified epochs of every target with \gls{DETECT}. We consider a detection to be real when the background-corrected magnitude in the difference image is brighter than the 80\% detection threshold. If not, we set the upper limit to equal the 80\% detection threshold.




\subsection{Summary of Targets with Significant Detections in Rubin \gls{DP1}}
\label{sec:results_detections}

In this Section we summarize the results for \todo{11} targets for which significant detections in Rubin \gls{DP1} were identified after analyzing them with \gls{DETECT}. These targets can be broadly separated into three groups, which we discuss in turn: (i) objects for which significant detections in Rubin \gls{DP1} were identified prior to the reported discovery date listed on \gls{TNS},  (ii) objects for which significant detections in Rubin \gls{DP1} were identified only after the previously reported discovery date, but for which pre-discovery upper limits from \gls{DETECT} are available, and (iii) objects for which no epochs we available in Rubin \gls{DP1} prior to the previously reported discovery date. For each target, we show a summary plot in Figure \ref{fig:detect_detections}, which shows the science image, template image and difference image of the highest SNR epoch of this target in the top row of each panel. The light curve of each target is shown in the bottom row of each panel.

\subsubsection{Targets with Significant Detections Prior to their Discovery Date}

\emph{AT 2024ahzi}: Figure \ref{fig:detect_detections}a shows the light curve of AT 2024ahzi. This transient was first reported by the \gls{YSE} collaboration \citep{murphey_2025} with a $g$-band magnitude of 22.242 mag on 19 Dec 2024. Unfortunately, the Rubin templates for the $i$-band were not usable (see Section \ref{sec:challenges}). However, epochs in other bands covered the time period between -31.0 and -10.0 days relative to the reported discovery date. Subsequent analysis with \gls{DETECT} confirmed two reliable $r$-band detections at -16.1 and -10.0 days relative to the discovery date, with $\textrm{m}_r$ = 23.04 and 22.68, respectively. It is unclear whether these pre-discovery epochs represent true precursor emission or the early phase of the transient, especially since the pre-discovery epochs can be interpreted as part of a smoothly rising light curve. Additional work on AT 2024ahzi will be crucial to disentangle these scenarios (e.g., see \citealt{de_soto_2026}).

\emph{AT 2025sfa}: The light curve for AT 2025sfa is shown in Figure \ref{fig:detect_detections}b. This target was discovered on \todo{27 July 2025} by \citet{oneill_2025} using data from the \gls{GOTO} telescope array \citep{steeghs_2022} with an $L$-band magnitude of 20.01 mag. Its discovery date is roughly 7 months after the end of \gls{DP1}, but we do have epochs ranging between -242.6 and -228.7 days relative to its discovery date. The target also appears to be in the center of its host galaxy. This usually makes finding injection locations difficult (see Section \ref{sec:challenges}), but similarly to AT 2024ahsx (see Section \ref{sec:results_nondetections}), there was a sufficiently nearby extended source that we could use to inject sources in instead. See Appendix \ref{app:suboptimal_inj_locs} for additional notes on using \gls{DETECT} in situations with suboptimal injection locations. Nevertheless, we find two significant detections in the $r$ and $g$ bands at MJD = 60652 (or at -228.7 days relative to the discovery date). The associated background-corrected magnitudes are $\textrm{m}_r$ = 22.87 mag and $\textrm{m}_g$ = 23.16 mag, which is 2.86 - 3.15 mag fainter than the reported discovery magnitude. Future work is necessary to confirm the nature of the transient itself and whether these pre-discovery detections could represent precursor emission. Alternatively, the central location of this target in its host galaxy is suggestive of a nuclear transient, such as a time-variable \gls{AGN} or a \gls{TDE}.

\emph{AT 2024ackk}: Figure \ref{fig:detect_detections}c shows the light curve of AT 2024ackk, first reported by \citet{tonry_2024b} using data from the \gls{ATLAS} on 27 November 2024 with a $c$-band magnitude of 19.29 mag. Rubin \gls{DP1} only contains two epochs for this target on -1.9 and +9.0 days relative to the discovery date. Further analysis with \gls{DETECT} confirms that both epochs show significant detections. The \gls{ATLAS} light curve also contains some detections in between. However, it seems that the source is also present in the Rubin template image. This implies that the reported magnitude from Rubin is likely underestimated and that the Rubin templates contain flux from the transient itself.

\emph{AT 2024aigl}: The light curve of the next target, AT 2024aigl, is shown in Figure \ref{fig:detect_detections}d. It was also originally found in Rubin \gls{DP1} \citep{anumarlapudi_2025} on \todo{9 December 2024} with a $z$-band magnitude of 22.4 mag. \gls{DP1} has additional epochs ranging between -21.0 and +2.0 days relative to the discovery date. Further analysis with \gls{DETECT} finds significant detections in the $r$, $i$, and $z$ bands in the epoch -20.1 days before the discovery date (at MJD = 60633). In contrast, we do not find significant detections in any other epochs, including the discovery date epoch. The 80\% detection threshold from \gls{DETECT} in the $z$-band for the discovery date epoch is equal to 22.2 mag, which excludes the original discovery magnitude of 22.4 mag. Additional work is needed to verify the nature of this transient and discern whether the pre-discovery epoch with significant detections in multiple bands contains true precursor emission.

\subsubsection{Targets with Significant Detection After their Discovery Date and Upper Limits Prior}

\emph{AT 2024ahzc}: The light curve for AT 2024ahzc is shown in Figure \ref{fig:detect_detections}e. This target was discovered with data from \gls{YSE} on 2 December 2024 with an $r$-band magnitude of 22.97 mag \citep{murphey_2025}. Rubin \gls{DP1} contains epochs between -22.9 and +8.0 days relative to the discovery date. However, we only find reliable detections with \gls{DETECT} between +2.0 and +8.0 days after its discovery date, which show a smoothly rising light curve. No significant pre-discovery emission was found. The deepest 80\% detection threshold from \gls{DETECT} is equal to 23.48 mag, which is 0.51 mag deeper than the \gls{YSE} discovery magnitude. 

\emph{AT 2024ahyy}: The next target is called AT 2024ahyy, and its light curve is shown in Figure \ref{fig:detect_detections}f. This target was also found using data from \gls{YSE} on 2 December 2024 with an $i$-band magnitude of 23.17 mag \citep{murphey_2025}. Similar to the previous target, Rubin DP1 contains epochs both before and after its discovery date, ranging between -23.5 and +8.4 days relative to the discovery date. We confirm reliable detections in the $r$-band between +3.4 and +6.4 days after its discovery date, which also show a smoothly rising light curve. However, no significant detections were found in the earlier epochs, and the deepest pre-discovery 80\% detection threshold from \gls{DETECT} is equal to 24.11 mag, almost 1 mag deeper than the discovery magnitude. However, the host galaxy was relatively small, which makes it a difficult target for \gls{DETECT}. No injection locations could be found in the same galaxy. Instead, injection locations were chosen in another nearby galaxy. This is not ideal, and the results of this target should be taken with a grain of salt. The quoted detections have \gls{SNR} between 7.7 and 10.8, and are likely robust. However, the quoted upper limits may be affected by the choice of injection locations. See Appendix \ref{app:suboptimal_inj_locs} for additional notes on suboptimal injection locations. 

\emph{AT 2024aigi}: Figure \ref{fig:detect_detections}g shows the light curve for AT 2024aigi, another target originally found in Rubin DP1 on 4 December 2024 \citep{freeburn_2025g}.\footnote{The TNS report of AT 2024aigi notes a discovery date of 4 December 2024. However, the authors of this paper cannot find a \gls{SNR}$>5$ detection at this date in Rubin \gls{DP1}. The discovery date on \gls{TNS} might be incorrect.} The other epochs in \gls{DP1} cover the period between -24.9 and +7 days relative to the discovery date. Interestingly, we only find one epoch with a significant discovery at +2.0 days relative to the discovery date (MJD = 60650.21). It is a very significant detection in the $g$-band, clearly visible in the difference image and with a background-corrected magnitude and \gls{SNR} of 21.1 mag and 78.4, respectively. It is 2.8 mag brighter that the 80\% detection threshold from \gls{DETECT}. However, the signal disappears completely in the same band at MJD = 60650.24, only $\sim$43 minutes later. We checked for the presence of asteroids near the coordinates of this target using \texttt{MPChecker}\footnote{https://www.minorplanetcenter.net/cgi-bin/checkmp.cgi} developed by the \gls{IAU}, but did not find any. Interestingly, \citet{malanchev_2025} also find a $g$-band detection with a very short timescale for another (unrelated) source in Rubin \gls{DP1}, similar to what is found here, and  concluded that their source was an M-dwarf flare. Further work is needed to confirm whether AT 2024aigi is also an M-dwarf flare, an instrumental artifact, a real event in the host galaxy with such a short timescale, or something else entirely.

\subsubsection{Targets with no Epochs Prior to their Discovery Date}

Here, we describe four targets for which no epochs were available in Rubin \gls{DP1} prior to the discovery date from TNS. Even though we restricted our sample to targets that were discovered \emph{after the start of \gls{DP1}} (see Section \ref{sec:sample}), this does not actually guarantee the presence of epochs before the discovery date. Some transients appeared in fields that were not covered until later in the \gls{DP1} campaign or were discovered with the first available epoch. While it is not possible to discover pre-SN variability for these targets, we still summarize the results of running \gls{DETECT} on their \gls{DP1} images here.

\begin{figure*}
    \centering
    \includegraphics[width=\textwidth]{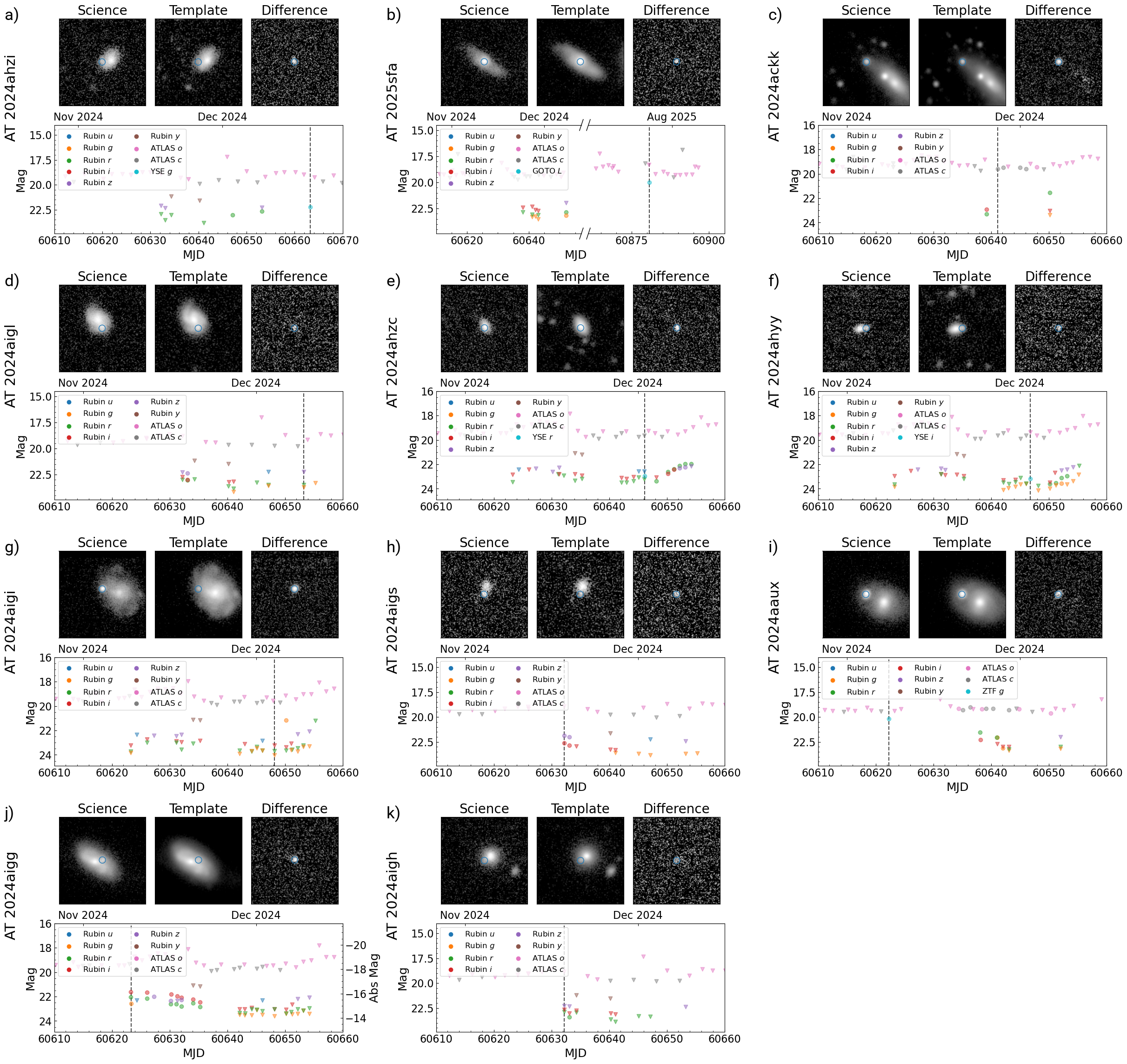}
    \caption{This figure shows the results from \gls{DETECT} of targets where we found at least one epoch with a significant detection. Within each panel, the top row shows the science image, template image and difference image of the highest \gls{SNR} epoch of this target. Each image cutout is 20 by 20 arcsec. The bottom row of each panel shows the light curve of that target. We include data from Rubin \gls{DP1} \citep{rubin_dp1_2025} that was analyzed with \gls{DETECT}. Circles denote epochs where the reported magnitude exceeded the 80\% detection threshold, whereas downward triangles represent non-detections, placed at the 80\% detection threshold. Note that we only show Rubin \gls{DP1} epochs when we were able to run \gls{DETECT}. This was not possible for every band and/or epoch due to incomplete templates, the target being too close to the edge of the exposure, or the inability to find suitable injection locations (see Section \ref{sec:challenges}). We also show the \gls{ATLAS} light curve for more context \citep{tonry_2018}. The vertical dashed line is placed at the discovery date of the target.}
    \label{fig:detect_detections}
\end{figure*}

\emph{AT 2024aigs}: Figure \ref{fig:detect_detections}h shows the light curve for AT 2024aigs. This target was found by \citet{freeburn_2025d} using Rubin \gls{DP1} on 18 November 2025\footnote{The TNS report of AT 2024aigs notes a discovery date of 19 November 2025, but subsequent follow-up revealed that it was actually visible in Rubin \gls{DP1} one day earlier. We use the updated discovery date in this work.} with an $r$-band magnitude of 22.02 mag. Interestingly, we could not find appropriate injection locations for the $r$-band template due to the small size of the host galaxy and the low quality of the template (see Section \ref{sec:challenges}). However, this was not a problem for the other bands. \gls{DP1} contains epochs ranging between 0 and +23.0 days relative to the discovery date. Further analysis with \gls{DETECT} confirms that the epochs directly surrounding the discovery date (between 0 and +0.9 days) show significant detections in the $i$ and $z$ bands. Epochs from +1.2 days onward do not have any significant detections. However, the 80\% detection threshold of the first non-detection is only 0.3 mag fainter than the earlier detection, which suggests that the transient has simply faded below the detection threshold. 

\emph{AT 2024aaux}: The light curve for AT 2024aaux is shown in Figure \ref{fig:detect_detections}i. It was discovered on 8 November 2024 using data from \gls{ZTF} by \citet{sollerman_2024} with a $g$-band magnitude of 20.16 mag. No epochs before its discovery date exist in Rubin DP1, but we do have epochs between +15.8 and +29.8 days after its discovery date. We confirm that multiple detections between +15.8 and +19.8 days after the discovery date are real. The light curve from \gls{ATLAS} also contains multiple detections during this same period. The Rubin detections show a smoothly declining flux, indicating Rubin \gls{DP1} caught the target as it was fading. However, the source itself is likely present in the template image, indicating that the reported magnitude of this target may also be underestimated. This also explains why the Rubin magnitudes are much fainter than the \gls{ATLAS} magnitudes.

\emph{AT 2024aigg}: The next target is AT 2024aigg \citep{freeburn_2025b}, which was originally discovered by Rubin DP1 on 9 November 2024 with an $i$-band magnitude of 21.48 mag. Its light curve is shown in Figure \ref{fig:detect_detections}j. Unfortunately, no epochs in \gls{DP1} exist before the discovery date, so we are not able to study its pre-discovery phase. However, we do have epochs between 0 and +30.0 days relative to the discovery date. Furthermore, we have significant detections in the $g$, $r$, $i$, and $z$ bands up to +11.9 days after the discovery date. The Rubin light curve shows a smoothly fading transient, which suggests that \gls{DP1} likely caught the tail end of this event. Since we know the luminosity distance ($D_L$ = 355.5 Mpc) of its host galaxy, we can convert the magnitudes to absolute magnitudes. The observed peak $r$-band absolute magnitude (which was found for the already fading transient) is equal to -15.7 mag, while the median 80\% detection threshold is equal to \todo{-14.73} mag.

\emph{AT 2024aigh}: The light curve of the next target, AT 2024aigh, is shown in Figure \ref{fig:detect_detections}k. This target was also first discovered in Rubin \gls{DP1} on 18 November 2024 with an $r$-band magnitude of 23.3 mag \citep{freeburn_2025c}. Similarly to before, there are no epochs in \gls{DP1} before its discovery date, but we do have epochs between 0 and +21.0 days relative to the discovery date. Analysis with \gls{DETECT} confirms one significant detection one day after the discovery date (+0.84 days) in the $r$-band. Interestingly, we do not find a significant detection at the discovery date epoch. The \gls{DETECT} 80\% detection threshold in the $r$-band is equal to 22.89 mag, which barely excludes the original discovery magnitude. Closer inspection reveals that the reported magnitude in the $r$ and $z$ bands between +0.0 and +0.8 days relative to the discovery dates are only slightly fainter than the 80\% detection threshold calculated by \gls{DETECT} (within $<$0.5 mag), although they would still not be classified as detections even if we used the 50\% detection threshold instead. Thus, while we currently cannot confidently say that these epochs are reliable detections, it is worth keeping in mind that in the future, when we have more accurate templates, these epochs might eventually be confirmed as real detections. 

\subsection{Summary of Targets with no Significant Detections in Rubin \gls{DP1}}
\label{sec:results_nondetections}

 Here, we cover the targets where we did not find any evidence for significant detections after using \gls{DETECT}. This is despite the fact that all of these targets had at least one epoch where forced photometry on the difference images exceeded the default Rubin \gls{SNR} $> 5$ threshold.

\begin{figure*}
    \centering
    \includegraphics[width=0.8\textwidth]{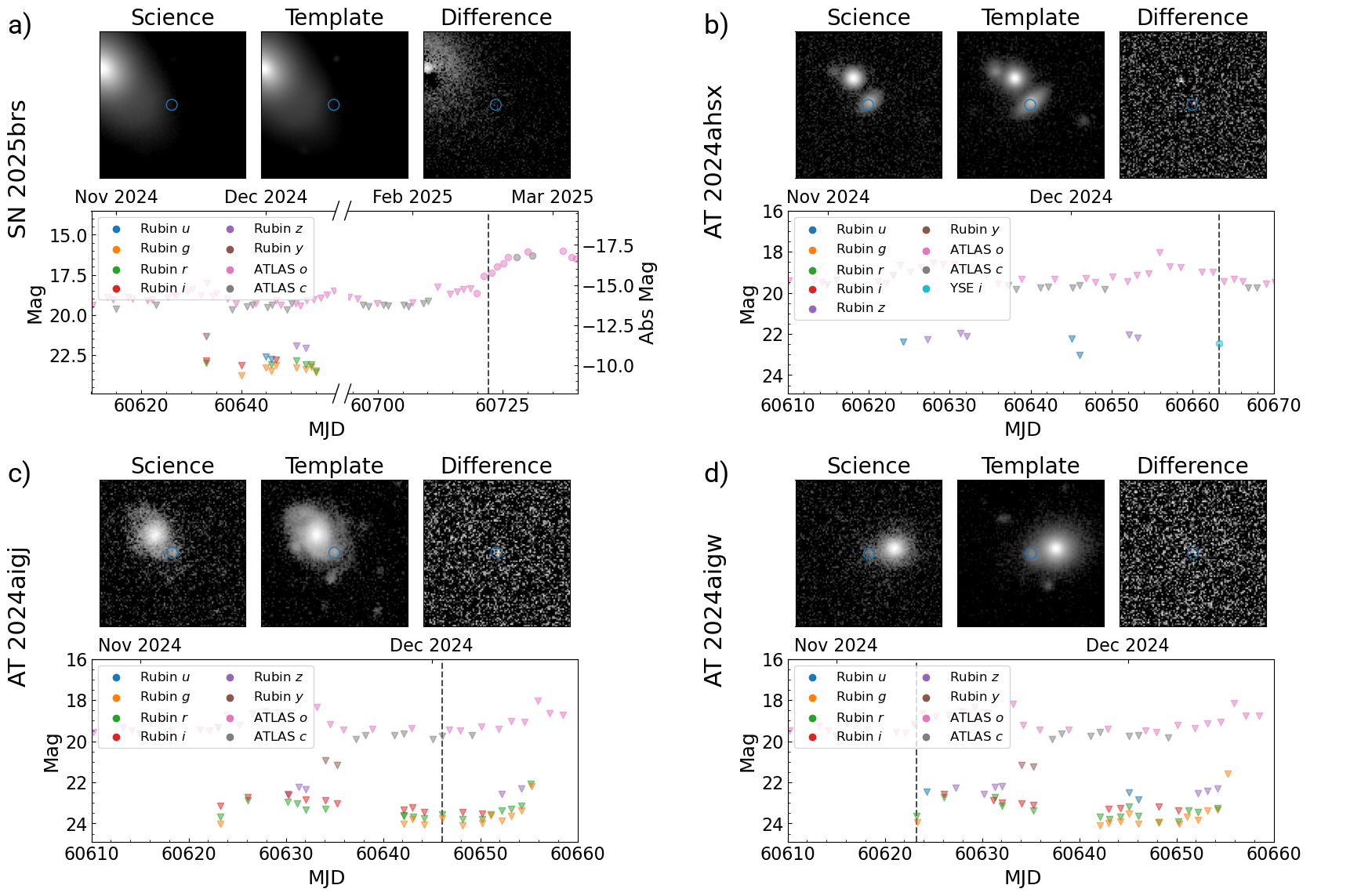}
    \caption{This figure shows the results from \gls{DETECT} of targets where we did not find any significant detections. Within each panel, the top row shows the science image, template image and difference image of the highest \gls{SNR} epoch of this target. Each image cutout is 20 by 20 arcsec. The bottom row of each panel shows the light curve of that target.  We include data from Rubin \gls{DP1} \citep{rubin_dp1_2025} that was analyzed with \gls{DETECT}. The downward triangles represent non-detections, placed at the 80\% detection threshold. Note that we only show Rubin \gls{DP1} epochs when we were able to run \gls{DETECT}. This was not possible for every band and/or epoch due to incomplete templates, the target being too close to the edge of the exposure, or the inability to find suitable injection locations (see Section \ref{sec:challenges}). We also show the \gls{ATLAS} light curve for more context \citep{tonry_2018}. The vertical dashed line is placed at the discovery date of the target.}
    \label{fig:detect_nondetections}
\end{figure*}

\emph{SN 2025brs}: The first target to discuss is SN \todo{2025brs}, shown in Figure \ref{fig:detect_nondetections}a. It is a \gls{SN Type Ia} discovered on \todo{16 February 2025} by \citet{rehemtulla_2025} using data from \gls{ZTF} with an $r$-band magnitude of 17.5 mag. This target was discovered after the end of \gls{DP1}, but we do have epochs between -89.2 and -67.2 days relative to the discovery date (or -104.0 and -82.1 days relative to the peak of the \gls{ATLAS} light curve). Interestingly, \todo{8} out of \todo{22} epochs in \gls{DP1} pass the default \gls{SNR} $>5$ threshold using fluxes obtained from forced photometry. We have already shown in Section \ref{sec:pipeline_description} that its MJD = \todo{60647} epoch is a false positive. However, subsequent analysis with \gls{DETECT} reveals that not a single significant detection was found in any of the Rubin \gls{DP1} bands. It is clear from the difference images alone that the image subtraction did not work as well as intended, likely due to the bright host galaxy of and SN \todo{2025brs}, and subsequent analysis with \gls{DETECT} confirmed these suspicions. However, we do now have reliable empirical upper limits for all these epochs. It is unsurprising, but worth pointing out that the upper limits obtained with \gls{DETECT} using data from Rubin \gls{DP1} are much deeper than the upper limits obtained from \gls{ATLAS}. The median upper limit on the magnitude obtained from \gls{ATLAS} during the period of Rubin \gls{DP1} is equal to \todo{19.05} mag (or \todo{-14.09} mag in absolute magnitude at the distance of SN 2025brs). In contrast, the median 80\% detection threshold from \gls{DETECT} using Rubin \gls{DP1} is equal to \todo{23.09} (or \todo{-10.05} mag in absolute magnitude), which is 7.09 mag deeper than the peak \gls{ATLAS} magnitude.

\emph{AT 2024ahsx}: We show the light curve for AT 2024ahsx in Figure \ref{fig:detect_nondetections}b. It was originally detected by \citet{murphey_2025} on 19 December 2024 using data from \gls{YSE} with an $i$-band magnitude of 22.45 mag. We have epochs ranging between -38.9 and -10.1 days relative to the discovery date. We do not find significant detections for this target using \gls{DETECT}. The 80\% detection threshold from \gls{DETECT} is similar to the reported discovery magnitude, with their difference ranging between -0.48 to +0.56 mag for the different epochs. However, it must be noted that it was not possible to find injection locations in the same host galaxy, since it is relatively small. Nevertheless, we were able to find injection locations in another nearby galaxy. However, we were still only able to find injection locations and run \gls{DETECT} on only the $u$ and $z$ bands. Refer to Appendix \ref{app:suboptimal_inj_locs} for additional notes on using \gls{DETECT} in cases with suboptimal injection locations.

\emph{AT 2024aigj}: The light curve for AT 2024aigj is shown in Figure \ref{fig:detect_nondetections}c. This target was first found in Rubin \gls{DP1} \citep{freeburn_2025f} on 2 December 2024. The $r$-band magnitude from Rubin \gls{DP1} at the discovery date is equal to 23.8 mag. We have additional epochs in Rubin \gls{DP1} that range between -22.8 and +9.2 days relative to the discovery date. Interestingly, we find no evidence for any significant detections in any bands. However, the reported magnitude in the difference image is only slightly fainter than the 80\% detection threshold found by \gls{DETECT} around the discovery date, especially in the $r$-band (the difference is $\lesssim$0.3 mag). However, these epochs would still be classified as non-detections even if we used the 50\% detection threshold instead. While we currently cannot claim that we have found any reliable detections, it is possible that this might change in the future once deeper template images are obtained.

\emph{AT 2024aigw}: Figure \ref{fig:detect_nondetections}d shows the light curve and triplet of AT 2024aigw, a target originally found in Rubin \gls{DP1} \citep{freeburn_2025} on \todo{9 November 2024}. We have epochs in \gls{DP1} that range between 0 and +31.9 days relative to the discovery date. However, subsequent analysis with \gls{DETECT} reveals that we find no evidence for any significant detections in any band. We note that this target was located only 2 pixels from the edge of its $i$-band science exposure at MJD = 60623, which possibly resulted in the false positive. This is a recurring issue that is covered in more detail in Section \ref{sec:challenges}.

\subsection{Common Challenges of Using \gls{DETECT}}
\label{sec:challenges}








\gls{DETECT} was developed to calculate reliable detection thresholds for targets embedded in bright nearby galaxies that possibly interfere with the default image subtraction pipelines. However, despite its robustness, there are a few common issues that can arise while using this pipeline. We outline several of these here.

\emph{Incomplete Template Images}: One issue to look out for are incomplete template images. For example, visual inspection of the \gls{DP1} $i$-band template of AT \todo{2024ahzi} shows that it is clearly incomplete, as the data cuts off in the middle of the exposure. This resulted in multiple $i$-band epochs being classified as detections by the default pipelines as well as \gls{DETECT}, but this is clearly due to the incorrect template. These bad templates are usually flagged with appropriate masks (e.g., \texttt{BAD} or \texttt{NO\_DATA}), and \gls{DETECT} provides functionality to quickly verify these masks. This will become less of an issue over time as the survey progresses and builds up reliable templates. However, we still strongly recommend to visually inspect each template that is used before continuing further analysis.

\emph{Overall Template Quality}: The overall quality of the templates used is also important. If better templates are used, the output of \gls{DETECT} will be more robust and the upper limits will be deeper. This becomes apparent after looking at the results presented by \citet{dong_2025}. They used template images obtained by \gls{DECam} to do image subtraction on corresponding science images from Rubin \gls{DP1} and were able to find 29 previously unreported transients hiding in the \gls{DP1} data. With these higher quality \gls{DECam} templates, they were also able to find detections in epochs in targets that went unnoticed by \gls{DETECT} (e.g., AT 2024aigj and AT 2024aigh). A similar approach was also used by \citet{freeburn_2025h}. This indicates that \gls{DETECT} will return more conservative limits (and potentially miss detections) when using shallow and low quality templates. However, it is important to note that, in this work, \gls{DETECT} is doing what it should based on the \gls{DP1} templates that were used. Fortunately, this issue will improve as the main Rubin \gls{LSST} survey begins and accumulates more and better templates over time. This will be especially important for targets with incomplete templates, such as the one discussed above, but also templates that contain flux of the target in question, such as AT 2024ackk and 2024auux (see Section \ref{sec:results_detections}).



\emph{Targets Near the Edge of the Exposure}: Another issue that can arise is when the target and host galaxy are near the edge of the science or template exposure. This can generate artifacts in the difference image following image subtraction, which subsequently can result in a false detection using the default \gls{SNR} $>$ 5 threshold. This happened for the $i$-band science exposure of AT \todo{2024aigw} at MJD = 60623. \gls{DETECT} was not designed to handle these cases, and we recommend to not run the pipeline on targets that border the edges of their science exposure (e.g., $\lesssim$ 40 pixels). We note that these problematic situations are also usually flagged in the associated image mask (with the \texttt{EDGE} keyword), but recommend to still visually inspect the science, template, and difference images for any obvious artifacts.





\emph{Finding Suitable Injection Locations}: Finally, we want to emphasize that the that results of \gls{DETECT} are conditional on finding appropriate injection locations. This will be straightforward for targets embedded in large nearby galaxies, which is the situation \gls{DETECT} was designed for. However, finding injection locations can be more difficult for smaller galaxies, or for targets located in the center of their host galaxies (e.g., nuclear transients). Nevertheless, usually you can still find injection locations by changing the default parameters (e.g., see Section \ref{sec:finding_injection_locations} and Appendix \ref{app:suboptimal_inj_locs}), but we recommend visually inspecting the resultant injection locations to make sure they are appropriate and to exercise caution when interpreting the results.

\subsection{Comparison of \gls{DETECT} and Rubin \gls{DP1} 5$\sigma$ Upper Limits}
\label{sec:comparison_upper_limits}

\begin{figure}
	\includegraphics[width=\columnwidth]{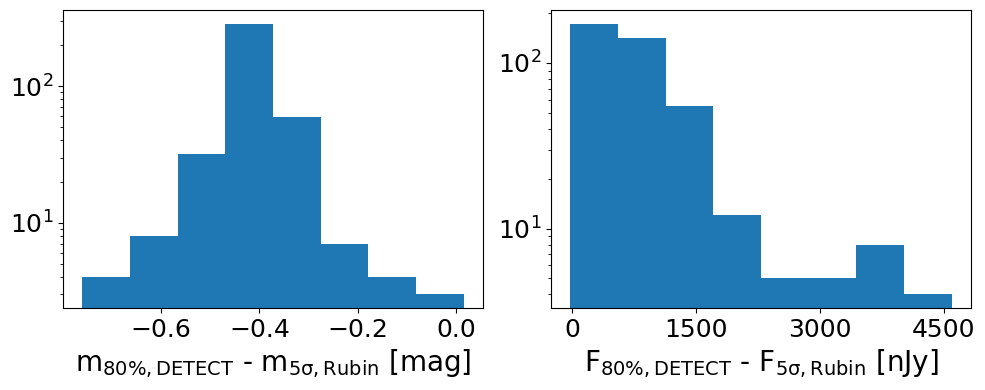}
    \caption{The left panel shows the difference in magnitude between the 80\% detection threshold obtained with \gls{DETECT} (m$_{\rm 80\%,\;DETECT}$) and the default 5$\sigma$ upper limit from Rubin (m$_{\rm 5\sigma,\;Rubin}$) for all epochs analyzed in this work (see Figures \ref{fig:detect_detections} and \ref{fig:detect_nondetections}). This sample contains targets that are located on top of bright host galaxies (see Section \ref{sec:sample}). The right panel shows the the difference in flux (F$_{\rm 80\%,\;DETECT}$ - F$_{\rm 5\sigma,\;Rubin}$).}
    \label{fig:delta_detect_rubin}
\end{figure}

We compare the 80\% detection threshold obtained with \gls{DETECT} (m$_{\rm 80\%,\;DETECT}$) to the default Rubin 5$\sigma$ upper limits (m$_{\rm 5\sigma,\;Rubin}$) in Figure \ref{fig:delta_detect_rubin} for all the epochs analysed in this work and shown in Figures \ref{fig:detect_detections} and \ref{fig:detect_nondetections}. We emphasize that m$_{\rm 5\sigma,\;Rubin}$ is not the expected ideal 5$\sigma$ point source depth of the whole survey, but rather the individual 5$\sigma$ threshold that is obtained after performing forced photometry on the difference image at the specific location of the target. It is clear that the upper magnitude limit obtained from \gls{DETECT} is lower than the default Rubin upper limit (i.e. m$_{\rm 80\%,\;DETECT}$ $<$ m$_{\rm 5\sigma,\;Rubin}$; or F$_{\rm 80\%,\;DETECT}$ $>$ F$_{\rm 5\sigma,\;Rubin}$). The reverse is true for only \todo{1} case. The median of m$_{\rm 80\%,\;DETECT}$-m$_{\rm 5\sigma,\;Rubin}$ is \todo{-0.4} mag (or $662$ nJy). As noted in Section \ref{sec:challenges}, the results of \gls{DETECT}, including the values of the upper limits, depend on the completeness and quality of the templates (though the value of m$_{\rm 5\sigma,\;Rubin}$ will also change with the quality of the templates). As the main Rubin \gls{LSST} survey begins, the quality of the template images will improve over time, which implies that the exact value of the upper limits will also change. 

\begin{figure}
    \centering
    \includegraphics[width=0.7\columnwidth]{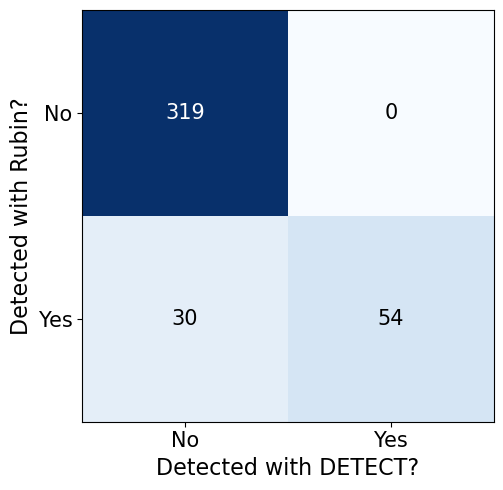}
    \caption{This confusion matrix shows whether a target is counted as a detection using the default Rubin 5$\sigma$ threshold and compares that to whether it is counted as a detection using the \gls{DETECT} 80\% detection threshold (i.e., the magnitude where 80\% of injected test sources were recovered at $\geq5\sigma$). We find that $\sim\todo{36}$\% of epochs that are registered as detections using the default Rubin 5$\sigma$ threshold are actually non-detections after analysis with \gls{DETECT}. However, keep in mind that the targets in this work were selected to be located in bright host galaxies that can possibly confuse the image subtraction pipelines. The false positive rate across all default 5$\sigma$ detections for the whole survey will not be this high.}
    \label{fig:cm}
\end{figure}

The confusion matrix shown in Figure \ref{fig:cm} tells a similar story. Here, we compare whether a target is counted as a detection using the default Rubin 5$\sigma$ threshold and compare that to whether it is counted as a detection using the \gls{DETECT} 80\% detection threshold. It is clear that all non-detections (using the default Rubin 5$\sigma$ threshold) stay non-detections after analysis with \gls{DETECT}. In contrast, a significant number of targets registered as detections using the default Rubin 5$\sigma$ threshold are in fact reclassified as non-detections after analysis with \gls{DETECT}. For the targets analyzed in this paper, only \todo{30} out of \todo{84} (or \todo{$\sim$36\%}) of epochs originally flagged as detections using the default Rubin 5$\sigma$ thresholds remain detections after analysis with \gls{DETECT}. A lot of these are cases where the target was embedded in a bright nearby galaxy. The mean background flux (see Section \ref{sec:background}) for cases that were registered as a detection using the default Rubin 5$\sigma$ threshold, but subsequently analysis with \gls{DETECT} revealed no detection, is equal to \todo{604} nJy, with the 90$^{\rm th}$ percentile equaling \todo{1555} nJy. Meanwhile, the distribution of the background flux for cases that are counted as detections using both methods is closer to 0; the mean is \todo{102} nJy and the 90$^{\rm th}$ percentile is equal to \todo{368} nJy. In short, if a target was registered as a non-detection using the default Rubin 5$\sigma$ threshold, it is unlikely that this will change after analyzing the target in more detail using \gls{DETECT}. However, the reverse is true. Targets registered as detections using the default Rubin 5$\sigma$ threshold are quite possibly false positives, especially if they are embedded in a bright nearby galaxy. Subsequent analysis with \gls{DETECT} is advised to make sure that the detection is real.

\begin{figure}
	\includegraphics[width=\columnwidth]{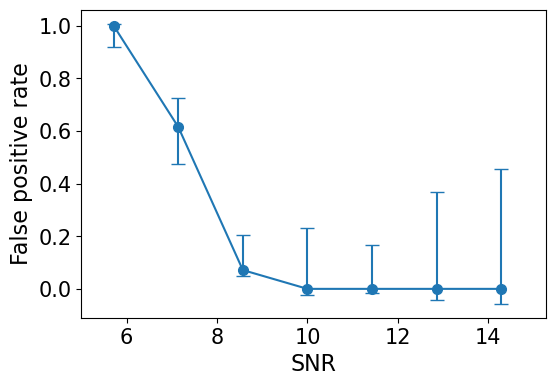}
    \caption{The false positive rate is plotted against the \gls{SNR} obtained by performing forced photometry at the location of the target. Epochs are considered false positives when they are counted as a detection using the default Rubin 5$\sigma$ threshold, but subsequently analysis with \gls{DETECT} reveals that no significant detection is found. This shows that most of the false positives occur when the \gls{SNR} is between 5 and 10, while no false positives are found when the \gls{SNR} $>$10. The error bars represent statistical uncertainties calculated with the beta distribution quantile technique \citep{cameron_2011}.} 
    \label{fig:snr_curve}
\end{figure}

The false positive rate was analyzed in greater detail in Figure \ref{fig:snr_curve}, where it is plotted against the \gls{SNR} obtained from performing forced photometry at the location of the target. It is clear that most of the false positives occur when the \gls{SNR} is between 5 and 10, while essentially no false positives in this work are found when the \gls{SNR} was greater than 10. However, keep in mind that the general shape of this curve will depend on the sample and on the quality of the templates used.

It is worth noting that the targets in this work are all selected to be embedded in bright nearby host galaxies that might confuse the default image subtraction pipeline. The false positive rate across all default 5$\sigma$ detections for the whole survey will likely not be this high, especially when as quality of templates improves over time. Additionally, we note that the \texttt{DiaSource} table contains the \texttt{reliability} column, which was still being calibrated and validated for \gls{DP1} \citep{rubin_dp1_diasource_table}. However, in future releases, this column will also be useful for identifying and filtering out false positives. Nevertheless, the results presented above are consistent with the results obtained with simulated data shown in Section \ref{sec:illustration_problem}, but now with real observations. This highlights the importance of using \gls{DETECT}, as the default Rubin 5$\sigma$ upper limits can be misleading in some cases. 

\section{Conclusion}
\label{sec:conclusion}

We developed \gls{DETECT}, a pipeline designed to calculate robust detection thresholds and upper limits in data from the Rubin Observatory. This is especially useful in cases where default image subtraction pipelines struggle, such as targets embedded in bright host galaxies, which is often the case when studying pre-SN variability. \gls{DETECT} works by performing a series of  source injection, image subtraction, and forced photometry for a wide range of magnitudes and injection locations.


The pipeline was first tested on simulated images from Rubin \gls{DP0}. We found that \gls{DETECT} correctly recovers sources up to $\textrm{m}_{\textrm{star}} \approx 23-24 \; \textrm{mag}$ when the host galaxy is relatively faint ($\textrm{m}_{\textrm{gal}} \gtrsim 16 \; \textrm{mag}$), as expected. However, when the host galaxy is bright ($\textrm{m}_{\textrm{gal}} \lesssim 15 \; \textrm{mag}$), this becomes more difficult. The bright host galaxy will interfere with the default image subtraction pipelines, and using the default \gls{SNR} $>$ 5 threshold can lead to false positives. Analysis with \gls{DETECT} shows that the 80\% detection threshold can vary with almost $\sim$\todo{2} mag, depending on the brightness of the host galaxy. This highlights the need for tools that can correctly handle these scenarios, such as \gls{DETECT}. Nevertheless, Rubin \gls{LSST} is expected to be a powerful probe of pre-SN variability and outbursts. To illustrate this, we have shown that Rubin, in combination with \gls{DETECT}, will be able to find the precursor outbursts of SN SN2009ip-like events up to $\sim130$ Mpc.

We then applied the pipeline to a sample of \todo{15} targets from Rubin \gls{DP1}. This demonstrated the robustness of the pipeline and its versatility in various use-cases. For example, we found reliable detections of a target before its original discovery date (AT \todo{2024ahzi}), although it is unclear whether these detections represent precursor emission or the early phase of the transient. We also verified that certain detections made with the default \gls{SNR} $>$ 5 threshold are indeed real (e.g., AT \todo{2024ahzc}) and we identified targets that had epochs incorrectly identified as detections (e.g., SN \todo{2025brs}). We calculated reliable empirical upper limits that can be used to constrain the underlying physical mechanisms (e.g., AT \todo{2024aigg}). While we initially set out to find pre-SN variability, we also found a potential nuclear transient (AT \todo{2025sfa}) and M-dwarf flare (AT \todo{2024aigj}), which shows the versatility of \gls{DETECT}, though more work is needed to confirm both these cases.

We show that the upper limits from \gls{DETECT} are typically \todo{$\sim$0.4} mag shallower than the default $5\sigma$ upper limits from Rubin. We also estimate that \todo{$\sim36$\%} of detections in this sample identified with the default ${\rm SNR} > 5$ threshold in the difference images may be spurious. However, most of the false positives occur when the \gls{SNR} is between 5 and 10, while no false positives were found when the \gls{SNR} was greater than 10. While the targets used in this study were deliberately chosen to be embedded in host galaxies where the default image subtraction pipelines will struggle, these results highlight the need for tools designed to handle such cases, like \gls{DETECT}.

We do note that the results of \gls{DETECT} are dependent on finding enough appropriate injection locations, which tends to be more challenging for smaller host galaxies that are further away. The quality of the templates that are used is also very important, as the output of \gls{DETECT} will be more robust and the resultant upper limits will be deeper if better templates are used. Furthermore, we recommend to always inspect the template images that are used to make sure that they are not incomplete, as well as being wary of the results obtained from targets that are located near the edge of their science exposure. 

The upcoming Rubin \gls{LSST} survey will allow us to study the rate of precursor emission across SN types and constrain the physical mechanisms that are responsible for the presence of the dense \gls{CSM} around SN Type IIn and SN with IIn-like features. Incorporating \gls{DETECT} in these analyses will be crucial, as we want to make sure any detected precursor emission is real and not affected by the brightness of its host galaxy. Lastly, while \gls{DETECT} was developed to study precursor emission, it is useful in any context where reliable upper limits are needed and detections are uncertain.


\begin{acknowledgments}
The authors would like to thank Melissa Graham for their insightful comments and helpful input during the development of the pipeline.

T.G. is a Canadian Rubin Fellow at the Dunlap Institute. The Dunlap Institute is funded through an endowment established by the David Dunlap family and the University of Toronto.

M.R.D. acknowledges support from the NSERC through grant RGPIN-2025-06224, the Canada Research Chairs Program, and the Dunlap Institute at the University of Toronto.

W.J.-G.\ is supported by NASA through Hubble Fellowship grant HSTHF2-51558.001-A awarded by the Space Telescope Science Institute, which is operated for NASA by the Association of Universities for Research in Astronomy, Inc., under contract NAS5-26555.

C.D.K. gratefully acknowledges support from the NSF through AST-2432037, the HST Guest Observer Program through HST-SNAP-17070 and HST-GO-17706, and from JWST Archival Research through JWST-AR-6241 and JWST-AR-5441.

\end{acknowledgments}

\begin{contribution}


\end{contribution}

%


\software{\texttt{Astropy} \citep{astropy_2013, astropy_2018, astropy_2022}, \texttt{Matplotlib} \citep{matplotlib_2007}, \texttt{NumPy} \citep{numpy_2020}, \texttt{SciPy} \citep{scipy_2020}, \texttt{DETECT}} 



\appendix

\section{Finetuning \texttt{n\_injections} and \texttt{n\_mag\_steps}}
\label{app:different_settings}

\begin{figure}
    \includegraphics[width=\columnwidth]{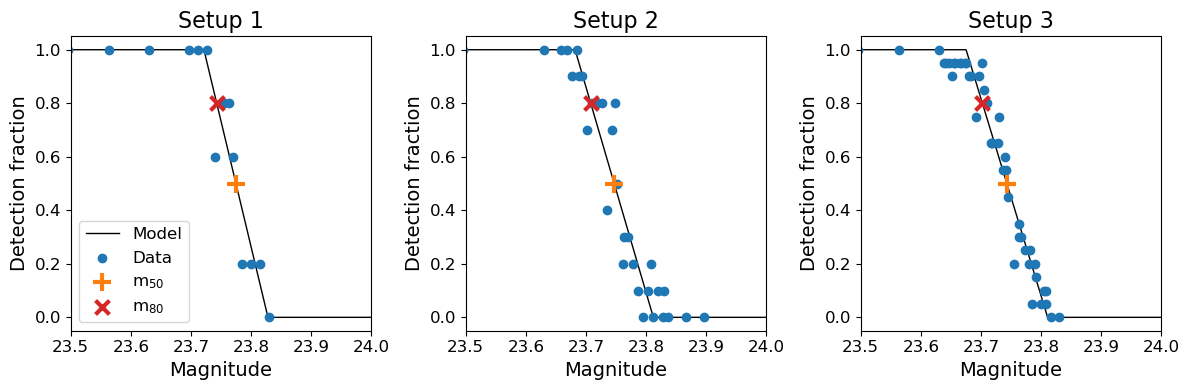}
    \caption{The recovery curve obtained after running \gls{DETECT} with different settings (see Table \ref{tab:setups}. We increasingly sample both axes in greater detail as we move from Setup 1 (left panel), to Setup 2 (middle panel), and finally Setup 3 (right panel). The resultant recovery curve model and detection thresholds become more reliable when we sample in greater detail.}
    \label{fig:different_setups_recovery_curves}
\end{figure}

As noted in Section \ref{sec:workflow}, \gls{DETECT} works by injecting fake sources into a number of injection locations (\texttt{n\_injections}) and calculating what fraction of those injected sources are detected. This is repeated for a wide range of magnitudes, and it is especially important to properly sample the transition region where the detection fraction goes from 1 to 0. This is done by first sampling magnitudes around the initial reported magnitude of the target in question. Afterwards, we keep honing in around the transition region and sample in greater detail. This is defined by the \texttt{n\_mag\_steps} parameter. For example, if \texttt{n\_mag\_steps = [4, 10]}, then we will first sample 4 magnitudes around the original reported magnitude of the target. \gls{DETECT} will try to identify the transition region, and then sample 10 magnitudes in the transition region in the next iteration. Note that you can effectively think of \texttt{n\_injections} and \texttt{n\_mag\_steps} as the resolution at which you sample both axes of the recovery curve. As expected, having enough injection locations and properly sampling the full magnitude range is crucial to obtaining a recovery curve and detection thresholds that are reliable. The default values of \gls{DETECT} work well in most cases (\texttt{n\_injections = 10} and \texttt{n\_mag\_steps = [4, 4, 4, 8, 20]}), but it advised to spend some time to make sure they are appropriate for your use-case. We provide some guidance to how to select these parameters by visualizing different setups, defined in Table \ref{tab:setups}. We use \gls{DETECT} with these different settings on AT 2024ackk (see Table \ref{tab:sample}), and the associated recovery curves are visualized in Figure \ref{fig:different_setups_recovery_curves}. The recovery curve of Setup 1 seems reasonable, but it is clear that both the detection fraction and magnitude range are not sampled in much detail. It is possible that the recovery curve model is fine, but it would be best to increase both parameters. In contrast, Setup 2 (which corresponds to the default settings of \gls{DETECT}) provides more detail and both the detection fraction and magnitude range are sufficiently sampled. The recovery curve model clearly follows the data. Finally, in Setup 3, we sample both axes in even greater depth. In general, higher values for \texttt{n\_injections} and \texttt{n\_mag\_steps} is better, at the cost of compute time.

\begin{figure}
    \includegraphics[width=0.7\columnwidth]{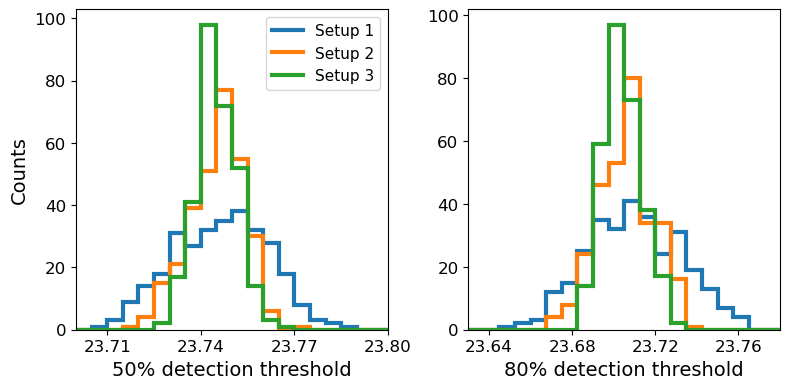}
    \centering
    \caption{The distribution of the 50\% detection threshold (left panel) and 80\% detection threshold (right panel) after running \gls{DETECT} \todo{300} times with the different settings defined in Table \ref{tab:setups}.}
    \label{fig:repeat}
\end{figure}

In Figure \ref{fig:repeat}, we estimate how much the output of \gls{DETECT} varies based on which settings are used. This is done by running \gls{DETECT} with the different settings presented in Table \ref{tab:setups} on AT 2024ackk \todo{300} times. We then fit the recovery curve and calculate the 50\% detection threshold (left panel of Figure \ref{fig:repeat}) and 80\% detection threshold (right panel of Figure \ref{fig:repeat}) for each iteration. It is clear that higher values for \texttt{n\_injections} and \texttt{n\_mag\_steps} result in less variability in the output. The standard deviation on the 80\% detection threshold of all iterations decreases significantly when sampling in greater detail; it goes from \todo{0.023} mag using Setup 1, to less than half that, \todo{0.009} mag, when using Setup 3. One can in principle repeat this process for any target to obtain uncertainties on the upper limits that \gls{DETECT} quotes, however this will take a lot of compute time. 

In summary, if the user samples the magnitude space sufficiently, and is able to locate enough possible injection locations, the uncertainty on the upper limits measured by \gls{DETECT} is low. We recommend to use the default values (i.e. \texttt{n\_injections = 10} and $\texttt{n\_mag\_steps = [4,4,4,8,20]}$), although more detail is always better and it is recommended to approach this on a case-by-case basis.

\begin{table}
\centering
\caption{An overview of the different values of \texttt{n\_injections} and \texttt{n\_mag\_steps} used to test the output of \gls{DETECT}. Setup 2 corresponds to the default values of \gls{DETECT}. We also show the standard deviation of the 50\% detection threshold ($\sigma_{50\%}$) and 80\% detection threshold ($\sigma_{80\%}$) after running \gls{DETECT} \todo{300} times with these settings.}
\label{tab:setups}
\begin{tabular}{c c c c c}
\hline
Name & \texttt{n\_injections} & \texttt{n\_mag\_steps} & $\sigma_{50\%}$ [mag] & $\sigma_{80\%}$ [mag]\\
\hline
Setup 1 & 5 & [4,4,4,8] & 0.015 & 0.023\\
Setup 2 & 10 & [4,4,4,8,20] & 0.009 & 0.014\\
Setup 3 & 20 & [4,4,4,8,15,20] & 0.006 & 0.009\\
\hline
\end{tabular}
\end{table}

\section{Finding injection locations for smaller galaxies}
\label{app:suboptimal_inj_locs}

\begin{figure}
    \includegraphics[width=\columnwidth]{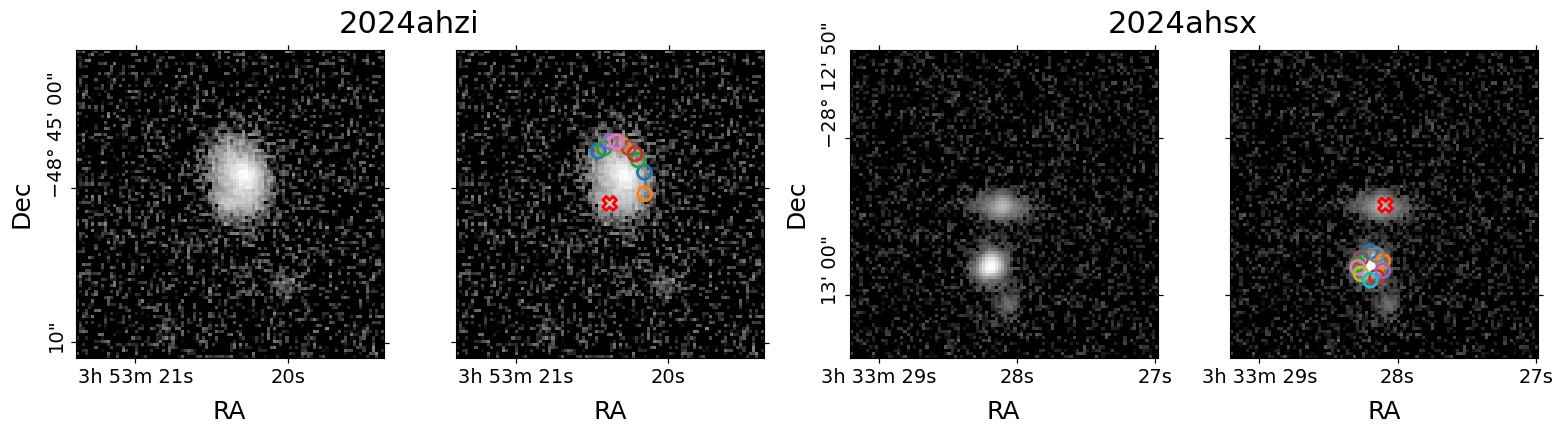}
    \caption{The left panel shows the $r$-band science exposure of AT \todo{2024ahzi} and the possible injection locations found with \gls{DETECT}, while the right panel shows the $z$-band for AT \todo{2024ahsx}. Finding injection locations is easier for nearby galaxies (e.g., SN \todo{2025brs}, see Figure \ref{fig:injection_locs}), although it is still possible to do for galaxies with smaller projected areas, provided that caution is used.}
    \label{fig:suboptimal_inj_locs}
\end{figure}

In Section \ref{sec:pipeline_description}, we demonstrated the \gls{DETECT} workflow using SN \todo{2025brs}, a \gls{SN Type Ia} embedded in a bright nearby galaxy. While this makes it an ideal target for \gls{DETECT}, it could still be useful to run this pipeline on targets that are further away and embedded in fainter galaxies. While it is certainly possible to do that, some caution is warranted while identifying possible injection locations. This is demonstrated using AT \todo{2024ahzi}, a target detected at $\alpha = 03^{h}53^{m}20.41^{s}$, $\delta = -48^{\circ}45'01.09"$ of unknown type \citep{murphey_2025}. Its redshift, at the time of writing, is also unknown, and the projected area on the sky of its host galaxy is markedly smaller than the example used in Section \ref{sec:pipeline_description}. The candidate injection locations found by \gls{DETECT} are visualized in the left panel of Figure \ref{fig:suboptimal_inj_locs}. As you can see, it is still possible to find suitable injection locations for galaxies with smaller projected areas. However, many of these locations are not in the same injection iteration, and will therefore not be injected together. This will cause the pipeline to take significantly more time to run.

Another interesting example is AT \todo{2024ahsx}, a target with coordinates $\alpha = 03^{h}33^{m}28.07^{s}$, $\delta = -28^{\circ}12'54.36"$ \citep{murphey_2025}. It appeared in the central regions of its small host galaxy, which makes finding appropriate injection locations difficult. However, there is another galaxy located nearby the host of AT \todo{2024ahsx}, and, as shown in the right panel of Figure \ref{fig:suboptimal_inj_locs}, the default settings of \gls{DETECT} have selected injection locations in that second galaxy. This might not be the behavior that was expected. However, these injection locations should still suffice for our purposes. We are simply looking for nearby locations with similar flux to our target and we assume that areas with similar flux will be affected similarly by the image subtraction pipeline, regardless of whether they happened to be located in the same galaxy. If this behavior is not desired, it is possible to reduce the the maximum search distance from the target (\texttt{max\_dist}) to avoid such cases, at the risk of not finding enough suitable injection locations within that range. 

It is most important to use \gls{DETECT} for targets embedded nearby bright galaxies, and identifying appropriate injection locations will be easier in these situations. However, it is still possible to use \gls{DETECT} for smaller galaxies, provided that caution is used when identifying injection locations. It is recommended to approach this on a case-by-case basis and visually inspect each target to be sure that you are content with the injection locations selected by \gls{DETECT}, and alter the settings accordingly.


\bibliography{bibtex}{}
\bibliographystyle{aasjournal}



\end{document}